\documentclass[fleqn,usenatbib]{mnras}
\usepackage[left]{lineno}

\usepackage{newtxtext,newtxmath}

\usepackage[T1]{fontenc}

\DeclareRobustCommand{\VAN}[3]{#2}
\let\VANthebibliography\thebibliography
\def\thebibliography{\DeclareRobustCommand{\VAN}[3]{##3}\VANthebibliography}

\usepackage{graphicx}	
\usepackage{amsmath}	
\usepackage{multirow}
\usepackage{lscape}
\usepackage{color}
\usepackage{lineno}

\usepackage{here}

\title[A disc-outflow system in G59.783+0.065]{A kinematic study of the disc-outflow system around a high-mass protostar G59.783+0.065 probed by methanol and water masers}

 \author[M. Nakamura et al.]{M. Nakamura$^{1 \textcolor{blue}{\star}}$, K. Motogi$^{2,3}$\thanks{E-mail:kmotogi@yamaguchi-u.ac.jp(KM); nakamura.momotaro@oshima-k.ac.jp(MN)}, H. Nakamura$^{2}$, Y. Yonekura$^{4}$, and K. Fujisawa$^{3}$ \\
$^{1}$Information Science \& Technology Department, National Institute of Technology, Oshima College, 1091-1 Komatsu Suo-Oshima, Oshima, Yamaguchi 742-2193, Japan\\
$^{2}$Graduate School of Sciences and Technology for Innovation, Yamaguchi University, 1677-1 Yoshida, Yamaguchi, Yamaguchi 753-8512, Japan\\
$^{3}$The Research Institute for Time Studies, Yamaguchi University, 1677-1 Yoshida, Yamaguchi, Yamaguchi
753-8511, Japan\\
$^{4}$Center for Astronomy, Ibaraki University, 2-1-1 Bunkyo, Mito, Ibaraki 310-8512, Japan
}

\date{Accepted 2023 September 6. Received 2023 August 31; in original form 2023 June 3}

\pubyear{2023}

\begin{document}

\label{firstpage}
\pagerange{\pageref{firstpage}--\pageref{lastpage}}
\maketitle

\begin{abstract}
Class II CH$_3$OH masers are used as a convenient tracer of disc-like structures in high-mass star formation. 
However, more than half of them show a complex distribution in Very Long Baseline Interferometry (VLBI) maps. 
The origin of such a complex distribution is still unknown. 
We conducted VLBI monitoring observations to unveil the origin of a complex class II CH$_3$OH maser in the high-mass star-forming region G59.783+0.065. 
We observed the CH$_3$OH maser at 6.7 GHz and the H$_2$O maser at 22 GHz to probe detailed circumstellar kinematics and structures by the Japanese VLBI network and the VLBI Exploration of Radio Astrometry. 
We found similar bipolar distributions in both masers, specifically two clusters located 2000 au apart along the East-West direction. 
We detected a linear distribution of CH$_3$OH masers in the Western cluster. 
A position-velocity diagram shows that the Western CH$_3$OH masers trace a rotating disc-wind or infalling component inside an edge-on disc-like structure. In contrast to the simple bipolar expanding motions of the H$_2$O masers, the CH$_3$OH masers exhibited complex motions despite their spatial coincidence. 
Some of the Eastern CH$_3$OH masers showed bipolar expansions similar to the H$_2$O masers, while others displayed random or even inward motions. 
Such complex kinematics and their close association with the H$_2$O maser could occur at the boundary between outflow and inflow. 
We suggest that the complex distribution of class II CH$_3$OH masers, like G59.783+0.065 arises from several distinct circumstellar structures that simultaneously achieve maser excitation. 
\end{abstract}




\section{Introduction}

Growing evidence suggests that high-mass stars can form via mass accretion through a disc-like structure, thanks to the remarkable progress of high angular resolution observations in the last two decades \citep[e.g.,][]{Cesaroni2007prpl, Cesaroni2017A&A, Beltran2016A&ARv}. 

Several important factors influence the stellar initial mass function (IMF), including the core mass function, outflow efficiency, and radiation feedback. 
The ultimate determination of stellar mass arises from disc fragmentation and subsequent mass accretion onto protobinary or multiples. 
Interferometric observations with resolutions high enough to image the circumstellar regions around distant young stellar objects are essential to study these processes and improve our knowledge of high-mass star formation \citep{Zhao2020SSRv}. 

Recent observations have resolved several infalling envelopes and disc-like structures around a high-mass protostar by using the Atacama Large Millimeter/submillimeter Array (ALMA) \citep[e.g.,][]{Ilee2018ApJL, Motogi2019, Zhang2019ApJ, Johnston2020A&A, Tanaka2020ApJ, Sanna2021A&A}. 
These observations have already indicated variations in the physical properties of disc-like structures, including their size, age, gravitational stability, and more. However, the current sample size remains limited, making it challenging to ascertain whether these diversities arise from initial conditions or evolutionary processes. 

Some interstellar masers are a powerful and convenient tool to investigate the three-dimensional structure and kinematics of disc-like structure and outflow in high-mass star-forming regions by using Very Long Baseline Intereferometry (VLBI). 
Class II CH$_3$OH maser at 6.7 GHz (hereafter, we call it simply `CH$_3$OH maser' in this paper) is excited closer to high-mass protostars (100--1000 au) compared to, for instance, Class I CH3OH masers. Therefore, it is a tracer that can be potentially used to infer the stellar location, study the circumstellar kinematics, and its interaction with outflow gas ejected along the polar cavities. 
A growing number of studies reported internal proper motion measured by long-term VLBI monitoring \citep[][etc.]{Sanna2010A&A...517A..71S, Sanna2010A&A...517A..78S, Goddi2011A&A, Matsumoto2011PASJ, Moscadelli2011, Moscadelli2013A&A, Sawada2013PASJ, Sugiyama2014A&A, Sugi2016,Bart2020}.

\begin{table*}
\caption{Summary of the VERA observations at 22 GHz. }
\begin{tabular}{lcccc}  
\hline\noalign{\vskip3pt} 
Epoch (\#) & 1   & 2   & 3  & 4    \\ \noalign{\vskip1pt} 
Date (UT) & 2016 Feb. 10  & 2016 Mar. 18  & 2016 Apr. 18  & 2016 May. 24  \\
Duration (day)  & 0  & 37    &  68   &  104   \\
\hline\noalign{\vskip3pt} 
Tracking centre $(\alpha, \delta)_\mathrm{J2000.0}$  & \multicolumn{4}{c}{$\mathrm{19^{h} 43^{m} \ 11^{s}.5, +23\degr 43\arcmin \  54\arcsec}$}\\
Antennas & \multicolumn{4}{c}{MIZ, IRK, OGA, ISG}  \\
Frequency$^{\rm a}$ (MHz) & 22227--22243 & 22228--22244 & 22228--22244  & 22228--22244 \\
Spectral channels for the maser   & 1024  & 1024 & 1024 &  1024  \\
Channel spacing (km s$^{-1}$)    & 0.21  & 0.21  & 0.21  & 0.21  \\
Image RMS 1$\sigma$ (Jy beam$^{-1}$)& 0.122 & 0.264 & 0.087 & 0.308 \\
Beam size (mas$^2$)  &  1.25 × 0.75 & 1.25 × 0.75 & 1.25 × 0.73 & 1.26 × 0.74  \\
Calibrators  & \multicolumn{4}{c}{3C345 (for fringe finder, bandpass), \  J1931+2243 (for phase)}  \\
\hline\noalign{\vskip3pt} 
\multicolumn{5}{l}{$^{\rm a}$ The frequency was changed at 2nd epoch for adjusting a Doppler correction caused by the earth orbital motion. }
\end{tabular}\label{VERA_obs}
\end{table*}

\begin{table*}
\caption{Summary of the JVN observations at 6.7 GHz.}
\begin{tabular}{lcccccc}  
\hline\noalign{\vskip3pt} 
Epoch (\#) & 1   & 2   & 3  & 4   & 5   & 6    \\ \noalign{\vskip1pt} 
Date (UT) & 2016 Aug. 25  & 2016 Oct. 31  & 2017 Jan. 22  & 2018 Nov. 22  & 2019 Jan. 11  & 2019 Apr. 30  \\
Duration (day)  & 0  & 67    & 150    & 819     & 869     & 978   \\
\hline\noalign{\vskip3pt} 
Tracking centre $(\alpha, \delta)_\mathrm{J2000.0}$ & \multicolumn{6}{c}{$\mathrm{19^{h} 43^{m} \ 11^{s}.25, +23\degr 44\arcmin \  03\arcsec.3}$}\\
Antennas & \multicolumn{6}{c}{MIZ, IRK, OGA, ISG, HIT, YMG$^{\rm a}$}  \\
Frequency (MHz) & 6667--6669   & 6667--6669   & 6667--6669   & 6667--6671  & 6667--6671  & 6667--6671 \\
Spectral channels for the maser$^{\rm b}$ & 512    & 1024 & 1024 & 4096 & 4096 & 4096   \\
Channel spacing (km s$^{-1}$)    & 0.176    & 0.088 & 0.088  & 0.044 & 0.044 & 0.044     \\
Image RMS 1$\sigma$ (Jy beam$^{-1}$)& 0.02  & 0.07   & 0.15   &  0.35  &  0.71  &  0.34   \\
Beam size (mas$^2$)  &  4.76 × 3.09 & 3.94 × 2.09 & 4.96 × 3.24 & 4.99 × 3.08 & 5.00 × 3.09 & 4.66 × 2.89 \\
Calibrators  & \multicolumn{6}{c}{3C454.3 (for fringe finder, bandpass), \  J1931+2243 (for phase)}  \\
\hline\noalign{\vskip3pt} 
\multicolumn{7}{l}{$^{\rm a}$No YMG data were obtained at the sixth epoch.}\\
\multicolumn{7}{l}{$^{\rm b}$ Spectral channels were increased by upgrades in the software correlator at the Mizusawa VLBI  Observatory.} 
\end{tabular}\label{JVN_obs}
\end{table*}

This CH$_3$OH maser shows several diverse distributions in VLBI maps, i.e., `ring (ellipse)', `arc', `linear', `pair' and `complex' \citep{Bart2009A&A}. 
The first three distributions were historically claimed to simply trace rotating disc-like structures or part of them only \cite[e.g.,][etc.]{Norris1993ApJ, Norris1998ApJ, Phillips1998MNRAS, Xu2001ChJAA}. Subsequently, more detailed observations showed complex scenarios combining either expanding and rotating motions, possibly associated with slow disc winds, or rotating and infalling motions \citep[e.g.,][]{Sugiyama2014A&A,Sugiyama2015PKAS, Sugi2016, Sanna2017A&A, Bart2018IAUS, Bart2020}. 
This fact indicates that the CH$_3$OH maser possibly traces more complex structure and kinematics near protostars than a previously considered. 

However, the number fraction of these three types is only half of the CH$_3$OH maser in statistical VLBI studies on 101 maser samples, which are compiled from the several VLBI surveys \citep[][]{Bart2009A&A, Bart2014, Bart2016A&A, Fujisawa2014PASJ}. 
Another half of the masers show a different distribution with respect to that expected in a disc-like structure, which can be otherwise classified as `pair' and `complex'. 
Their physical origins are still unclear, and some CH$_3$OH masers may be excited by a protostellar outflow \citep{Minier2000,Minier2001evn5,Moscadelli2011}.

We have performed multi-epoch VLBI monitoring of the complex CH$_3$OH maser sources in G59.783+0.065 (hereafter, G59). 
We have also observed the H$_2$O maser at 22 GHz (hereafter, we simply call it `H$_2$O maser'). 
We aimed to unveil the physical origin of the complex CH$_3$OH maser, comparing the internal proper motions of two maser species. 
High-mass star-forming region G59 is also known as IRAS19410+2336. 
The trigonometric parallaxes indicate that G59 is located at a distance of 2.16 kpc from the Sun \citep{xu2009}.  
The total infrared luminosity is $\sim 10^4 L_{\sun}$ \citep{Mart2008}, which corresponds to a (proto)stellar mass range of 10--20 $M_{\sun}$ based on a stellar evolutionary model under high accretion rates \citep[e.g.,][]{Hosokawa2009ApJ,Hosokawa2010ApJ}. 
\citet{Tanti2011} derived $\sim$ 10 $M_{\sun}$ through SED fitting. 

The systemic velocity of the natal molecular clump is measured to be $V_{\rm LSR}$ = 22.4 km s$^{-1}$ based on CH$_3$CN and H$_2$CO lines from the hot core \citep{Rod2012}. 
Here, $V_{\rm LSR}$ indicates the line-of-sight (LoS) velocity with respect to the local standard of rest (LSR). 
\citet{Carral1999} detected a pair of radio continuum sources at 8 GHz by the Very Large Array (VLA). 
The CH$_3$OH \citep{Caswell1995MNRAS1} and H$_2$O \citep{Lada1981ApJ} masers are associated with these centimetre continuum sources. Another class II CH$_3$OH maser at 12.2 GHz is also associated \citep{Caswell1995MNRAS2}. 

Two dust clumps were detected by the IRAM 30m telescope, one in the North and another in the South, at 1.2 mm \citep{Beuther2002A&A...383..892B}. 
The brighter southern clump was found to consist of four sub-clumps identified by the Plateau de Bure Interferometer (PdBI) at 2.6 mm \citep{Beu2003}, implying active cluster formation. 
\citet{Rod2012} further resolved the clump containing the CH$_3$OH and H$_2$O masers using PdBI at 3 mm and 1.4 mm. 
The total clump mass was estimated to be 187 $M_{\sun}$ at 3 mm. 
Both masers are associated with a further sub-core referred to as `13-s' at 1.4 mm, which has an estimated mass of 8 $M_{\sun}$ and likely represents an individual accretion envelope.

\begin{figure*}
	\includegraphics[width=0.96\hsize]{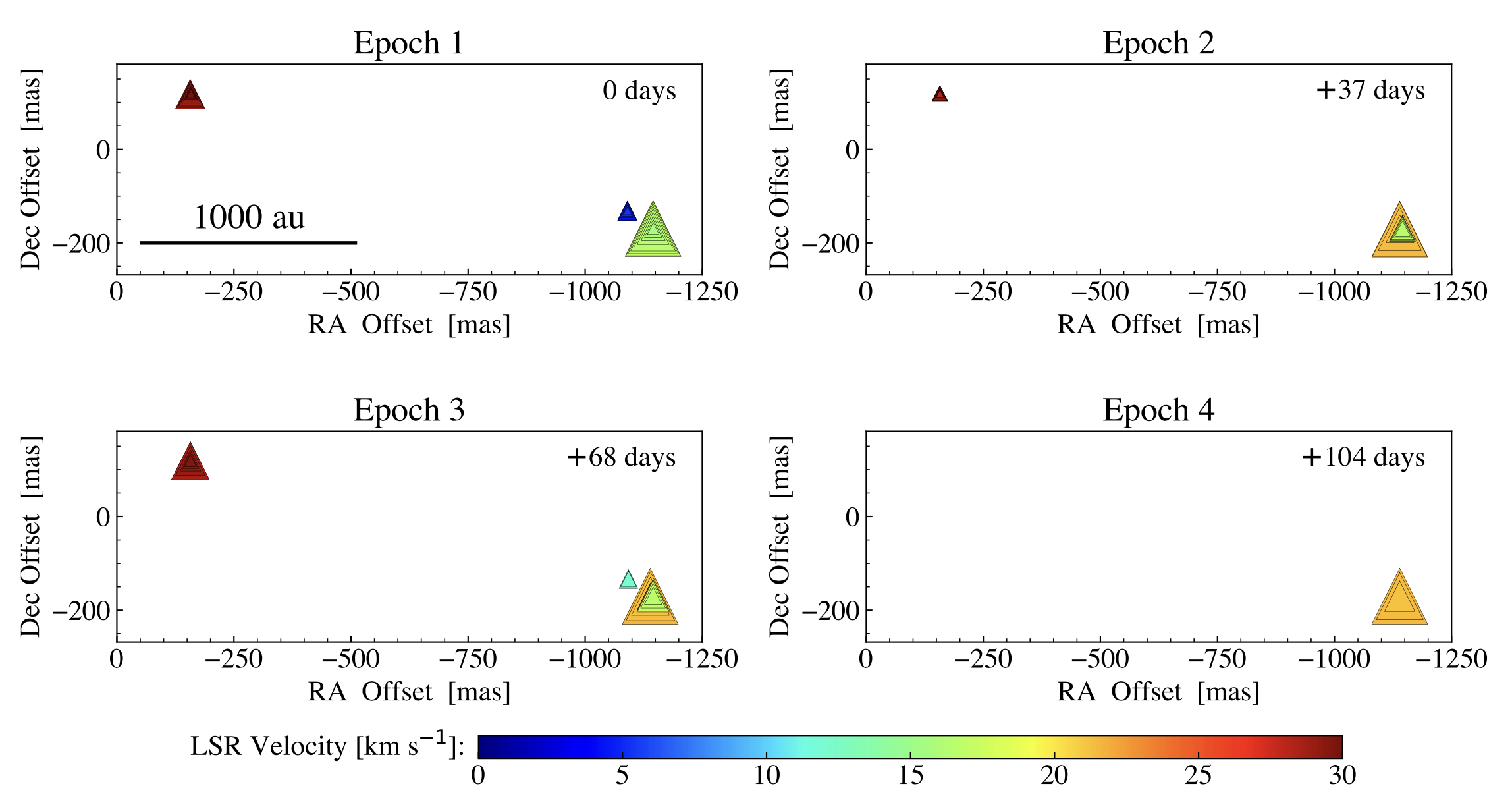}
    \caption{Spatial distributions of the H$_2$O masers in G59. 
    Each triangle shows the relative position of a maser spot, with the size of triangle proportional to the peak flux on a linear scale. The colour indicates the LSR velocity of the corresponding maser spot. The relative day from the first epoch is presented in the upper-right corner of each panel. 
    The coordinate origin at the first epoch is set to the position of the phase-referenced CH$_3$OH maser spot (see main text). }
    \label{fig:VERA_map_fig}
\end{figure*}
\begin{figure*}
	\includegraphics[width=0.8\hsize]{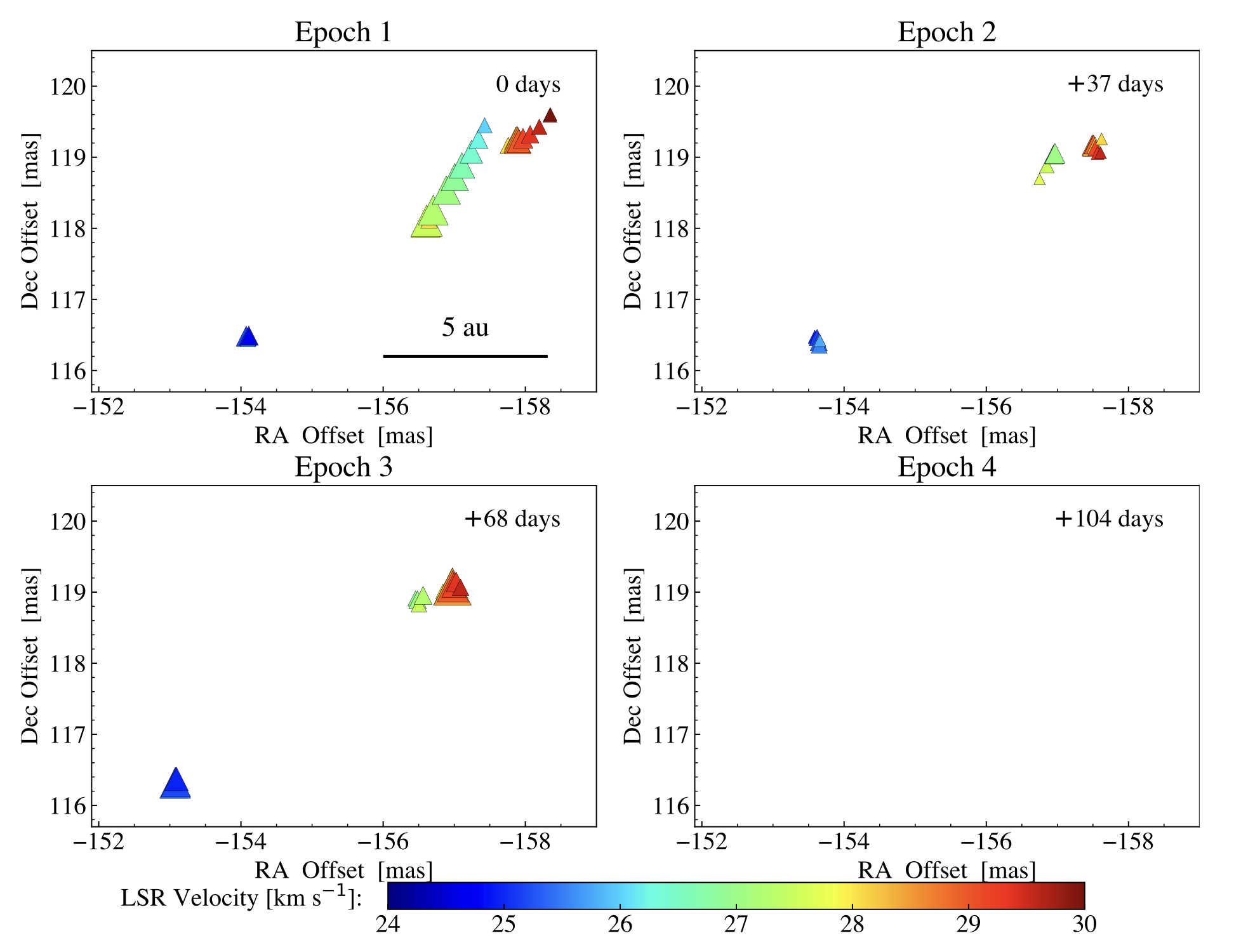}
    \caption{Zoomed-up view of the Eastern H$_2$O maser cluster in Figure \ref{fig:VERA_map_fig}. The marker sizes are proportional to the 0.7 power of the peak fluxes for highlighting fainter maser spots. The velocity range represented by the colour bar is narrower than Figure \ref{fig:VERA_map_fig}. 
    \label{fig:VERA_map_fig_zoom1}}
\end{figure*}
\begin{figure*}
	\includegraphics[width=0.85\hsize]{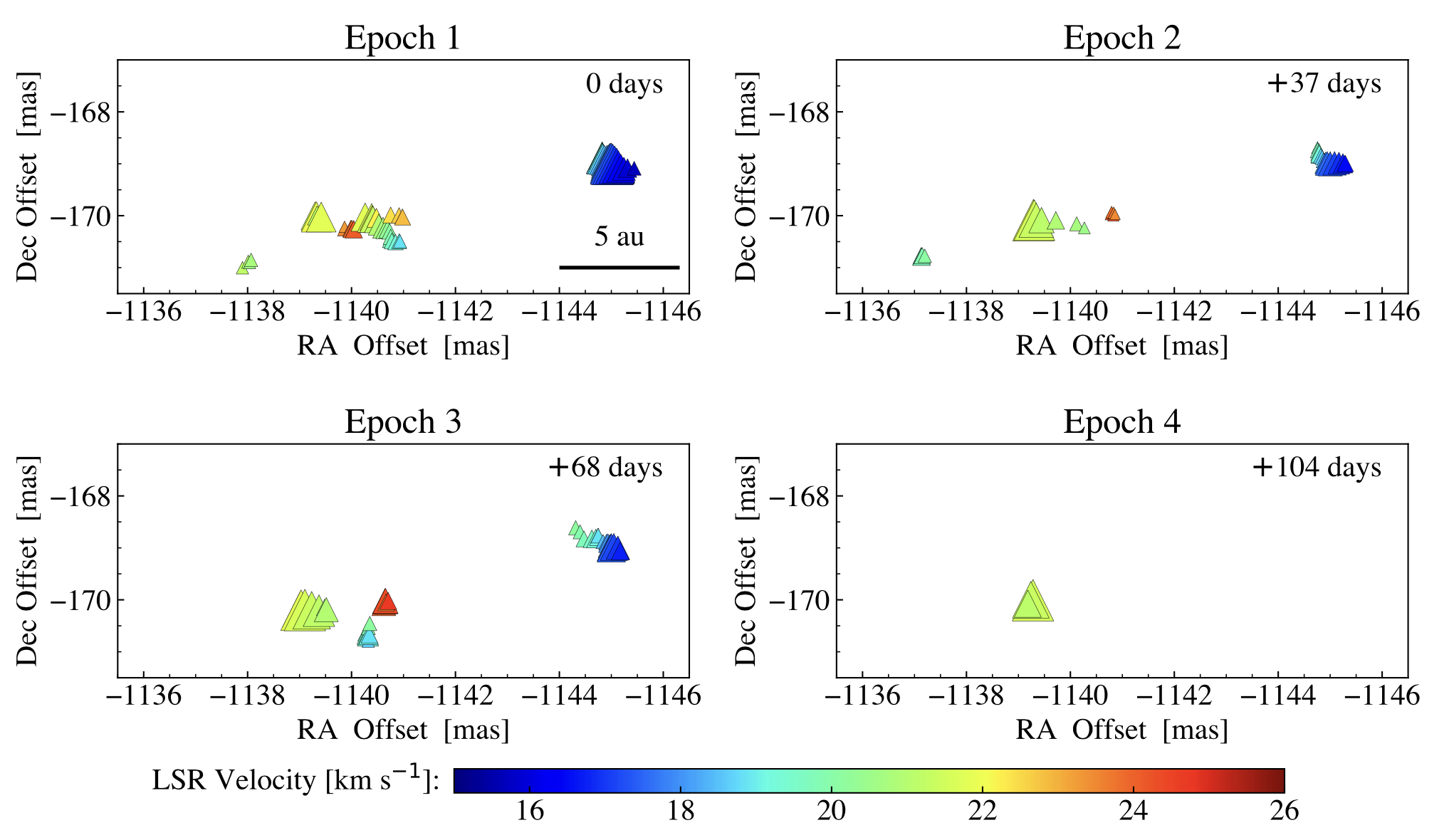}
    \caption{Same as Figure \ref{fig:VERA_map_fig_zoom1} but for the Western H$_2$O maser cluster. The velocity range represented by the colour bar is narrower than Figure \ref{fig:VERA_map_fig}. 
    \label{fig:VERA_map_fig_zoom2}}
\end{figure*}

\citet{Beu2003} detected a bipolar CO outflow around 13-s oriented in the East-West direction. 
\citet{Mart2008} detected a near-infrared (NIR) ro-vibrational H$_2$ line, which is a well-known outflow tracer, as well as NIR continuum emission scattered by the outflow cavity along the same NE-SW direction. 
This scattering emission was also observed at 3.6/4.5 $\mu$m in $Spitzer$ IRAC data \citep{Qiu2008ApJ}. 
We presents a summary of the circumstellar environment in G59 region in Appendix \ref{IR_view}. 

The LoS velocities ($V_{\rm LSR}$) of the CH$_3$OH masers range 14--27 km s$^{-1}$. 
This velocity width of 13 km s$^{-1}$ is among the group of sources with broader velocity widths found by \citet[][their Figure 5]{Green2017MNRAS}. 
The spectrum of the CH$_3$OH maser exhibits a double-peaked profile with peaks at 19 and 27 km s$^{-1}$.
The CH$_3$OH maser displayed sinusoidal-like variations before MJD 55650, where the relative amplitude varied factor of $\sim$ 2. On the other hand, the peak flux remained almost stable at around 20 Jy during the quiescent phase \citep[][]{Goedhart2004MNRAS,Szymczak2018MNRAS}. 

There are several interferometric images of class II CH$_3$OH masers. 
\citet{Minier2000} imaged 12 GHz CH$_3$OH masers using the Very Long Baseline Array (VLBA) in January 1999.  
The 6.7 GHz CH$_3$OH masers were observed twice, once by the Multi-Element Radio Linked Interferometer Network (MERLIN) on December 31, 2004 \citep[][]{Darwish2020} and again by the European VLBI Network (EVN) on March 13, 2010 \citep[][]{Bart2014}. 
A consistent `pair' of maser clusters, separated by $\sim$ 2000 au (900 milliarcseconds: mas) along the E-W direction, was observed in both the EVN and MERLIN data. We note that while the overall distribution is classified as a `pair', each maser cluster exhibits a `complex' distribution. 
\citet{Darwish2020} also conducted imaging observations of the H$_2$O maser by (e-)MERLIN. 
However, no reports of internal proper motion have been made for either the CH$_3$OH or H$_2$O masers. 

This paper reports new proper motion measurements for these masers, using the Japanese VLBI Nerwork (JVN) and the VLBI Exploration of Radio Astrometry (VERA). 
We have compared the kinematics of two maser species to study the origin of the complex CH$_3$OH maser. 
We also report our interpretation in terms of a disc-outflow system. 

\begin{figure*}
	\includegraphics[width=\hsize]{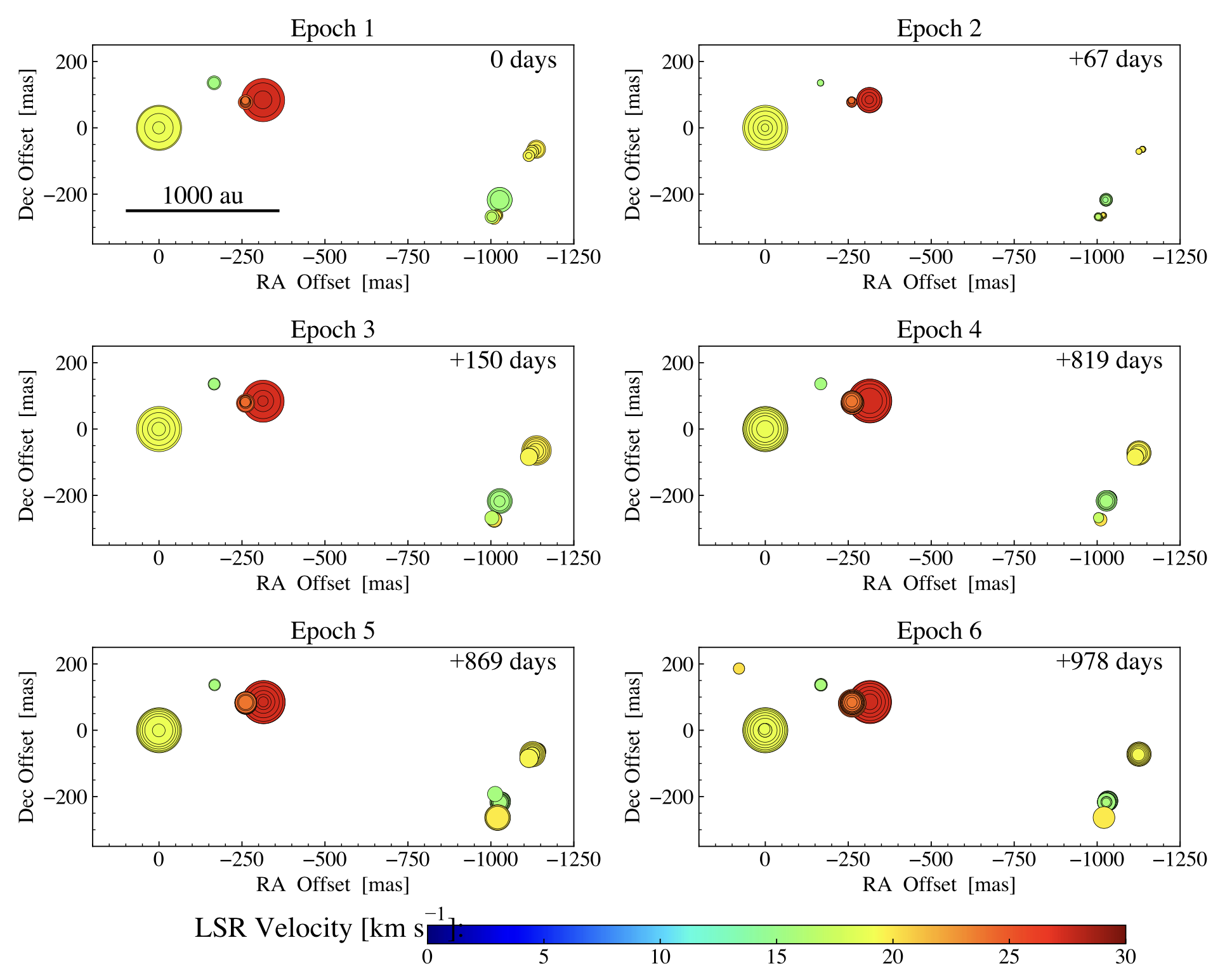}
    \caption{Same as Figure \ref{fig:VERA_map_fig} but for the CH$_3$OH masers. We used a circle instead of a triangle here. 
    The coordinate origin is the brightest CH$_3$OH maser spot in each epoch. The velocity range represented by the colour bar is the same as that in Figure \ref{fig:VERA_map_fig} for easy comparison.  
    The origins of Figure \ref{fig:VERA_map_fig} and Figure \ref{fig:JVN_map_fig} are only identical at the first epoch, based on the phase-referencing (see Section 2.3). For other epochs, the origins were different between the two figures since phase-referenced imaging was not successful for both masers. 
    }
    \label{fig:JVN_map_fig}
\end{figure*}

\section{Observations and data reduction}

\subsection{VERA observation}
The H$_2$O maser transition (${J_K}_{\rm a}$$_{K}$$_{\rm c}$ = $6_{16}-5_{23}$) at 22.23508 GHz was monitored monthly from February to May 2016 using VERA, which includes four VLBI stations: Mizsawa (MIZ), Iriki (IRK), Ogasawara (OGA), Ishigaki (ISG). 
Table \ref{VERA_obs} summarises our VERA observations.
There were four epochs during 104 days. 
The baseline length ranged from 1018 km (IRK-ISG) to 2270 km (MIZ-ISG), providing a beam size of 1 mas ($\sim$ 2 au at a distance of 2.16 kpc).  

All of the observations were conducted in the 1-beam mode. We scanned 3C345 10 min every 1.5--2.0 hours as a fringe finder and bandpass calibrator. 
We performed phase-referencing VLBI in switching mode, employing J1931+2243 as a phase calibrator that was 2$\degr$.89 apart from G59. 
The cycle time was set to 100 sec (30 sec for G59, 60 sec for J1931+2243, and 5 sec for slew time). 
The typical flux density of J1931+2243 was ranged from 0.3 to 0.4 Jy, providing sufficient brightness for accurate determination of interferometric delay, rate, and phase within 1-min integration. 
The total on-source time for G59 was about 75 min. 

Left-handed circular polarised (LHCP) signals were quantised by 2-bit sampling and recorded on hard disc devices at 1 Gbps. 
Correlation processing was carried out using the software correlator developed at the National Astronomical Observatory of Japan (NAOJ) Mizusawa campus.
We obtained a total 256-MHz wide bandwidth, which was divided into 16 IFs of 16 MHz each. One of these IFs was dedicated to the maser, while the remaining 15 IFs were used for calibrators. 
The single 16-MHz IF for the maser was divided into 1024 channels, providing a channel spacing of 0.21 km s$^{-1}$ for all epochs. 

The absolute position of the brightest H$_2$O maser spot of 17.36 km s$^{-1}$ at the first epoch was $(\alpha, \delta)_{\rm J2000.0}$ = $({\rm 19^{h}\ 43^{m} \ 11^{s}.18764, +23^{\degr} 44\arcmin \  03\arcsec.0094}$). 
Unfortunately, a phase-referenced image was obtained only at the first epoch. 
This is mainly due to insufficient atmospheric calibrations since the separation angle between J1931+2243 and G59 is larger compared to the usual cases of VERA astrometry in the dual beam system \citep{Honma2008PASJ}. 
The uncertainty in this position is better than $\sim$ 1 mas, taking into account errors in baseline length, thermal image noise, and atmospheric phase fluctuations \citep[e.g.,][]{Motogi2011MNRAS}. 
We note that this uncertainty does not include the error in absolute position of the phase calibrator. This could be a potential source of systematic error.

\begin{table*}
\caption{Identified maser features}
\scalebox{1.0}{
\begin{tabular}{ccrrrrrrrrrrr}
\hline
 Cluster$^{\rm a}$ & Name & $V_\mathrm{peak}$$^{\rm b}$ & $\Delta V$$^{\rm c}$ & $S_\mathrm{peak}$$^{\rm b}$ & $\Delta$RA$^{\rm d}$   & $\Delta$DEC$^{\rm d}$& \multicolumn{6}{c}{Number of maser spots in each epoch} \\
 &  & (km s$^{-1}$) &  (km s$^{-1}$) & (Jy) & (mas) & (mas)  & ep1 & ep2 & ep3 & ep4 & ep5 & ep6 \\
\hline
\multicolumn{13}{c}{H$_2$O maser features}\\ \noalign{\vskip1.5pt} 
E & w-1   & 25.15 & 0.84 & 2.05  & -154.090 (0.017) & 116.476 (0.017) & 5 & 5 & 5 & - &  &  \\
E & w-2   & 27.47 & 2.11 & 7.46  & -156.833 (0.302) & 118.451 (0.487) & 11 & 7 & 4 & - &  &  \\
E & w-3   & 28.73 & 1.90 & 4.64  & -157.942 (0.197) & 119.277 (0.146) & 10 & 8 & 9 & - &  &   \\
W & w-4   & 21.99 & 1.69 & 11.01 & -1139.334 (0.037) & -170.030 (0.054) & 9 & 11 & 9 & 4 &  &    \\
W & w-5   & 23.89 & 1.26 & 2.31  & -1139.992 (0.071) & -170.255 (0.019) & 7 & 6 & 7 & - &  &   \\
W & w-6$^{\rm e}$   & 17.36 & 4.21 & 27.03 & -1144.981 (0.177) & -168.987  (0.084) & 21 & 18 & 8 & - &  &   \\ \noalign{\vskip1pt} 
\multicolumn{13}{c}{Short-lived features not used for proper motion measurements.} \\ \noalign{\vskip1.5pt} 
W & w-a   & 4.09  & 2.53 & 3.26  & -1090.073  (0.094)  & -131.007 (0.076) & 13 & - & - & - &  &  \\
W & w-b   & 20.73 & 0.63 & 1.07  & -1138.007  (0.063)  & -170.908 (0.056) & 4 & 5 & - & - &  &  \\
W & w-c   & 21.57 & 4.42 & 9.51  & -1140.543  (0.212) & -170.140 (0.190)  & 22 & - & - & - &  &  \\
W & w-d   & 11.90 & 0.63 & 1.27  & -1092.522 (0.033) &  -133.430 (0.011)  & - & - & 4 & - &  &  \\
W & w-e   & 19.28 & 1.47 & 1.65  & -1140.316 (0.022)  & -170.651 (0.096)  & - & - & 8 & - &  &   \\
W & w-f   & 19.28 & 1.69 & 1.17  & -1144.630 (0.169)  & -168.793 (0.079)  & - & - & 9 & - &  &  \\
\hline 
\multicolumn{13}{c}{CH$_3$OH maser features}\\ \noalign{\vskip1.5pt} 
E & m-1$^{\rm e}$  & 19.25 & 1.06 & 19.01 & -0.121  (0.640) & -0.331 (0.600) & 7 & 12 & 9  & 19 & 19 & 21 \\ 
E & m-2  & 17.32 & 0.36 & 1.55  & -20.445 (0.140) & -6.197 (0.189) & 3 & 3  & 3  & 8  & 6  & 7 \\
E & m-3  & 17.14 & 0.53 & 2.13  & -25.529 (0.659) & -7.844 (0.236) & 4 & 3  & 3  & 5  & 8  & 9 \\
E & m-4  & 15.56 & 0.35 & 1.79  & -165.977 (0.168) & 135.722 (0.157) & 3 & 3  & 2  & 3  & 4  & 5  \\
E & m-5  & 24.70 & 1.23 & 1.97  & -260.561 (0.290) & 80.099 (2.385) & 8 & 13 & 11 & 26 & 27 & 32\\ 
E & m-6  & 27.15 & 1.05 & 17.41 & -313.786 (0.594) & 83.363 (1.220) & 7 & 10 & 9  & 18 & 18 & 18 \\
W & m-7  & 16.97 & 0.53 & 1.72  & -1003.316 (0.513) & -268.261 (0.368) & 4 & 8  & 2  & 2  & -  & -  \\
W & m-8  & 20.13 & 0.53 & 1.31  & -1009.536 (0.162) & -273.207 (0.053) & 4 & 5  & 3  & 4  & -  & -  \\
W & m-9  & 19.78 & 0.35 & 1.32  & -1018.565 (0.023) & -263.700 (0.471) & 3 & 6  & -  & -  & 13 & 13 \\
W & m-10 & 15.39 & 0.87 & 5.80  & -1026.592 (0.161) & -217.186 (0.288) & 6 & 12 & 8  & 13 & 14 & 14 \\
W & m-11 & 14.69 & 0.35 & 0.87  & -1029.893 (0.553) & -213.909 (0.542) & 3 & 5  & -  & 13 & 12 & 16  \\
W & m-12 & 19.43 & 0.53 & 1.23  & -1114.213 (0.214) & -83.831 (0.184)  & 4 & -  & 3  & 5  & 6  & -  \\
W & m-13 & 19.60 & 0.70 & 1.52  & -1125.450 (1.538) & -71.219 (1.108) & 5 & 5  & (1)$^{\rm f}$  & 12 & 12 & 14 \\
W & m-14 & 19.95 & 0.70 & 3.12  & -1136.994 (0.595) & -64.560 (0.754) & 5 & 5  & 8  & 6  & 7  & 2  \\ \noalign{\vskip1 pt} 
\multicolumn{13}{c}{Short-lived features not used for proper motion measurements.}\\ \noalign{\vskip1.5pt} 
W & m-a  & 15.39  & 0.53 & 0.69  & -1043.648 (1.006) & -222.300 (0.481) & 4 & - & - & - & - & - \\ 
E & m-b  & 15.55  & 0.09 & 10.15  &  -1013.129 (0.168) & -192.151 (0.058) & - & - & - & - & 3 & - \\
E & m-c  & 20.18  & 0.13 & 2.89  &   78.740 (0.312) & 185.986 (0.186) & - & - & - & - & - & 4 \\ 
\hline 
\multicolumn{13}{l}{$^{\rm a}$ E: Eastern cluster, W: Western cluster. }\\
\multicolumn{13}{l}{$^{\rm b}$ $V_{LSR}$ and flux density of the peak maser spot in each maser feature. }\\
\multicolumn{13}{l}{$^{\rm c}$ Total velocity ranges of spectral channels where maser spots were detected. }\\
\multicolumn{13}{l}{$^{\rm d}$ Relative coordinate of maser features (see main text). The parenthesis values indicate positional errors in mas. }\\
\multicolumn{13}{l}{$^{\rm e}$ The brightest maser features used for the phase-referencing and self-calibration.}\\
\multicolumn{13}{l}{$^{\rm f}$ Only single spot was detected for m-13 feature at the third epoch. This spot was ignored in any analysis in this paper. }\\
\end{tabular}
}
\label{Feature_VERA_JVN}
\end{table*}

\begin{table}
\caption{The barycentres for H$_2$O masers}
\begin{tabular}{lcccc}
\hline
Epoch & 1 & 2 & 3 & 4  \\ \hline
Relative \ RA (mas)   & -648.862  & -648.870  & -648.562 & - \\
Relative \ Dec (mas)  & -24.843  & -24.734  & -24.821 & - \\
\hline 
\multicolumn{5}{l}{Note: Barycentre cannot be calculated in the fourth epoch (see main text).}\\
\end{tabular}
\label{barycenter_H2O}
\end{table}

\subsection{JVN observation}
The CH$_3$OH maser transition at 6.668519 GHz ($J_K$ = $5_1-6_0$ $A^{+}$) has been monitored by the six JVN stations (Hitachi: HIT, Yamaguchi: YMG, and four VERA stations) from August 2016 to April 2019. 
Table \ref{JVN_obs} summarises the details of our JVN observations. 
There were a total of six epochs of observations conducted over 978 days. 
The observations were carried out approximately once every 3--4 months, except for the third and fourth epochs, which had a separation of around 22 months between them. 
No data were obtained in the YMG station at the sixth epoch because of instrumental trouble. 
The baseline lengths in these observations ranged from 293 km (IRK-YMG) to 2270 km (MIZ-ISG), providing a beam size of 3 mas ($\sim$ 6 au at 2.16 kpc).  

The data recording and correlation procedure for the JVN observations were the same as in the VERA observations, but at a higher data rate of 2 Gbps.
We used a single 512-MHz wide IF for the calibrators with coarse 1-MHz channel spacing. 
The detailed spectral setup for the maser varied depending on the observing epoch (see Table \ref{JVN_obs}). 
We used another 2 or 4-MHz narrow IF for the maser. 
The spectral channels were ranged from 512 to 4096, resulting in channel spacing of 0.044--0.176 km s$^{-1}$. 

We scanned 3C454.3 for 10--15 min as a fringe finder and bandpass calibrator. 
We performed phase-referencing VLBI in switching mode by using the phase calibrator J1931+2243 again. 
The cycle time was set to 180 sec (60 sec for each source and 30 sec for slew time). 
The total on-source time for G59 was about 120 min. 

We note that the phase-referencing for the CH$_3$OH maser also succeeded only at the first epoch. 
This is also attributed to the atmospheric conditions, particularly due to insufficient ionospheric corrections. 
The absolute position of the brightest CH$_3$OH maser of 19.25 km s$^{-1}$ at the first epoch was determined to be$ (\alpha, \delta)_\mathrm{J2000.0}$ = $(\mathrm{19^{h}\ 43^{m} \ 11^{s}.2710, +23\degr 44\arcmin \ 03\arcsec.178}$). 
The positional accuracy is also estimated to be better than 1 mas, similar to the case of 22 GHz maser. The positional error of phase calibrator was not taken into account again. 
This CH$_3$OH maser spot has offsets of (+1145 mas, +169 mas) relative to the phase-referenced H$_2$O maser spot. Considering the absolute positions of the referenced spots for both H$_2$O and CH$_3$OH masers, we present maps of their relative positions in this paper, with the referenced CH$_3$OH maser centred at the origin. 

\begin{table*}
\caption{The barycentres for CH$_3$OH masers}
\begin{tabular}{lcccccc}
\hline
Epoch & 1 & 2  & 3  & 4  & 5  & 6  \\
\hline
Relative \ RA (mas)  & -606.431 & -606.586 & -606.514 & -607.833 & -607.770 & -608.116 \\
Relative \ Dec (mas)  & -46.702  & -46.890   & -46.910  & -46.416  & -46.393  & -46.240  \\
\hline 
\end{tabular}
\label{barycenter_CH3OH}
\end{table*}

\begin{table*}
\caption{Internal proper motions}
\scalebox{1.0}{
\begin{tabular}{ccrrrrrrrrrrrr}
\hline
 Cluster$^{\rm a}$ & Name & $\mu_\alpha$ & $\sigma {\mu_\alpha}$& $\mu_\delta $ & $\sigma {\mu_\delta}$ & $V_\mathrm{RA}$ & $\sigma V_\mathrm{RA}$ & $V_\mathrm{DEC}$ & $\sigma V_\mathrm{DEC}$ & $V_\mathrm{prop}^{\rm a}$ & $\sigma V_\mathrm{prop}$ & $V_{\rm LoS}^{\rm b}$ \\
 &  & \multicolumn{2}{c}{(mas yr$^{-1}$)} & \multicolumn{2}{c}{(mas yr$^{-1}$)} & \multicolumn{2}{c}{(km s$^{-1}$)}  & \multicolumn{2}{c}{(km s$^{-1}$)} & \multicolumn{2}{c}{(km s$^{-1}$)} &(km s$^{-1}$) \\
\hline
\multicolumn{13}{c}{H$_2$O masers}\\ \noalign{\vskip1.5pt}  
E & w-1   & 3.840  & 0.324 & -1.068 & 0.300 & 39.3  & 3.3  & -10.9 & 3.1  & 40.8 & 4.5 & 2.75\\
E & w-2   & 1.092  & 0.588 & 0.372  & 1.188 & 11.2  & 6.0  & 3.8   & 12.2 & 11.8 & 13.6 & 5.07  \\
E & w-3  & 3.012  & 0.528 & -0.504 & 0.720 & 30.8  & 5.4  & -5.2  & 7.4  & 31.3 & 9.1  & 6.33  \\
W & w-4    & -0.516 & 0.204 & -0.888 & 0.360 & -5.3  & 2.1  & -9.1  & 3.7  & 10.5 & 4.2 & -0.41   \\
W & w-5   & -3.360 & 2.244 & 1.116  & 0.492 & -34.4 & 23.0 & 11.4  & 5.0  & 36.3 & 23.5 & 1.49 \\
W & w-6 & -2.040 & 0.960 & -0.108 & 0.276 & -20.9 & 9.8  & -1.1  & 2.8  & 20.9 & 10.2 & -5.04  \\ \noalign{\vskip1.5pt}  
\multicolumn{13}{c}{CH$_3$OH masers}\\ \noalign{\vskip1.5pt}  
E & m-1 & 0.554  & 0.022  & -0.238  & 0.056  & 5.671  & 0.230  & -2.442  & 0.578  & 6.174  & 0.622  & -3.15 \\
E & m-2 & 0.296  & 0.136  & 0.006  & 0.185  & 3.029  & 1.392  & 0.062  & 1.897  & 3.029  & 2.353  & -5.08 \\
E & m-3 & 0.242  & 0.099  & -0.098  & 0.084  & 2.480  & 1.017  & -1.004  & 0.863  & 2.675  & 1.333  & -5.26 \\
E & m-4 & -0.004  & 0.067  & 0.250  & 0.062  & -0.041  & 0.683  & 2.555  & 0.635  & 2.555  & 0.932  & -6.84 \\
E & m-5 & -0.030  & 0.062  & 0.141  & 0.048  & -0.306  & 0.636  & 1.439  & 0.494  & 1.471  & 0.805  & 2.30 \\
E & m-6 & -0.271  & 0.034  & 0.025  & 0.040  & -2.770  & 0.347  & 0.253  & 0.413  & 2.782  & 0.540  & 4.75 \\
W & m-7 & 0.254  & 0.277  & 0.367  & 0.117  & 2.599  & 2.835  & 3.760  & 1.197  & 4.571  & 3.077  & -5.43 \\
W & m-8 & 0.104  & 0.140  & -0.105  & 0.206  & 1.066  & 1.432  & -1.080  & 2.111  & 1.517  & 2.551  & -2.27 \\
W & m-9 & -0.321  & 0.024  & -0.091  & 0.044  & -3.283  & 0.245  & -0.930  & 0.446  & 3.412  & 0.509  & -2.62 \\
W & m-10 & -0.113  & 0.015  & 0.104  & 0.053  & -1.158  & 0.156  & 1.065  & 0.541  & 1.573  & 0.563  & -7.01 \\
W & m-11 & -0.159  & 0.106  & 0.035  & 0.080  & -1.628  & 1.083  & 0.361  & 0.824  & 1.667  & 1.360  & -7.71 \\
W & m-12 & -0.042  & 0.378  & -0.010  & 0.228  & -0.431  & 3.868  & -0.099  & 2.339  & 0.443  & 4.520  & -2.97 \\
W & m-13 & 0.523  & 0.141  & -0.646  & 0.106  & 5.356  & 1.446  & -6.613  & 1.085  & 8.510  & 1.808  & -2.80 \\
W & m-14 & -0.150  & 0.056  & -0.131  & 0.026  & -1.534  & 0.577  & -1.340  & 0.270  & 2.036  & 0.637  & -2.45 \\ \hline
\multicolumn{13}{l}{$^{\rm a}$ Total proper motions on the celestial plane.}\\
\multicolumn{13}{l}{$^{\rm b}$ Internal LoS velocities ($V_{\rm peak}-V_{\rm sys}$). }\\
\end{tabular}
} 
\label{proper_motion_VERA_JVN}
\end{table*}

\subsection{Data Calibration by AIPS}
We used the National Radio Astronomy Observatory (NRAO) Astronomical Image Processing System \citep[AIPS: ][]{Greisen2003ASSL} software package for standard calibration and synthesis imaging. 

The visibility amplitude was calibrated using the `ACCOR' and `APCAL' tasks at 22 GHz, with a measured system noise temperature ($T_{\rm sys}$) at each VERA station. 
At 6.7 GHz, $T_{\rm sys}$ was not available in the HIT and YMG stations. 
Therefore, we used the `ACFIT' task for amplitude calibration in the `Template method' instead of `APCAL'. 
We determined relative antenna gains by comparing the total power spectra of the maser. 
Then the flux was scaled, referencing $T_{\rm sys}$ at the IRK station. 
The fringe finders were used for calibrating clock offsets (`FRING') and bandpass responses (`BPASS'). 
We then solved residual delay, rate, and phase using J1931+2243 by `FRING' task. 
After applying Doppler correction using the `CVEL' task, the phase-referenced image for the brightest maser spot was deconvolved using the `IMAGR' task. 
We further calibrated the residual rate and phase using the `FRING' task with the peak maser channel in each epoch to improve image quality for measuring internal proper motions. 
Finally, the phase and amplitude were self-calibrated using the brightest maser spots, combining the `IMAGR' and `CALIB' tasks. 
The minimum solution interval was 15 sec for the H$_2$O maser and 30 sec for the CH$_3$OH maser. 
All imaging and deconvolution (CLEAN algorithm) were done adopting natural weighting.
We searched maser spots with a 7-$\sigma$ detection limit. 
Parameters of detected maser spots were extracted by elliptical Gaussian fitting with the `JMFIT' task. 

We identified a maser `feature', which is a physical gas clump emitting maser emission \citep[e.g., ][]{Sanna2010A&A...517A..71S} based on the following criteria: 
\begin{enumerate}
 \item A feature is a cluster of two or more maser spots within a few mas.  
 \item The spots comprising the feature appear in successive spectral channels.
 \item The same feature was detected in at least three observing epochs. 
\end{enumerate}
\ \ The position of an identified maser feature was given by the error-weighted average of the constituent maser spots. 
Each feature was named using the following convention: 
We first set `w' or `m’ depending on a maser species (w: water, m: methanol). 
Then we added a serial number if a maser feature satisfies all three criteria. 
Instead, an alphabet was given on a maser feature that does not satisfy the third criterion (i.e., it is not detected in at least three epochs). 
We did not use the latter `short-lived' features for proper motion measurements. 

In this paper, positions of H$_2$O maser spots (and features) are presented as follows: First, we estimated the relative positions of maser spots with respect to the brightest maser spot used for self-calibration in each observing epoch. Then, we subtracted the constant offset of (+1145 mas, +169 mas) as mentioned above. 

The positions of CH$_3$OH maser spots are simply given as relative positions with respect to the brightest CH$_3$OH maser spot in each epoch. The brightest CH$_3$OH maser spots always belong to the identical CH$_3$OH maser feature, including the phase-referenced spot at the first epoch. This fact means that the coordinate origins of both maser data matched at the first epoch, where the phase-referenced imaging was succeeded for both masers. Although the positions of H$_2$O masers after the second epoch are not directly linked to the relative coordinate of CH$_3$OH masers, it is still accurate enough to compare the overall distributions of two maser species. 


\section{Results}
\subsection{Masers distributions}

\begin{figure*}
	\includegraphics[width=\hsize]{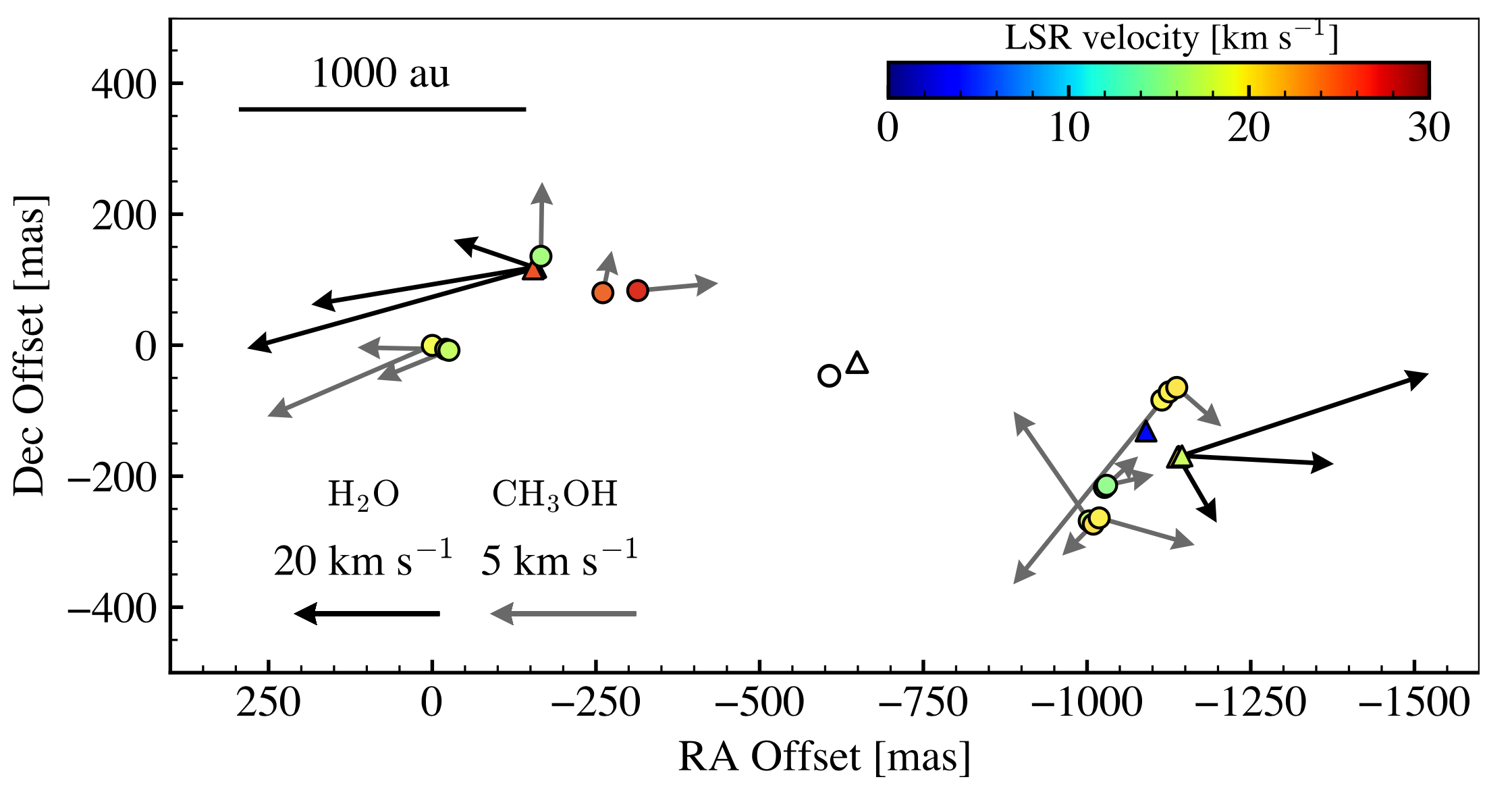}
    \caption{Internal proper motions of the CH$_3$OH masers (grey vectors) and the H$_2$O masers (black vectors). 
    Each maser feature is indicated by a circle for CH$_3$OH and a triangle for H$_2$O masers. 
    Here, we only showed locations of long-lived maser features at the first epoch. 
    The coordinate origin for both maser species is set to the position of the phase-referenced CH$_3$OH maser spot, i.e., $(\alpha, \delta)_\mathrm{J2000.0}$ = $(\mathrm{19^{h}\ 43^{m} \ 11^{s}.2710, +23\degr 44\arcmin \ 03\arcsec.178}$). 
    The colour of each marker represents the LSR velocity of the maser feature. The barycentres, which are the referenced position of the proper motions, are shown by open markers. 
    The length of each vector is proportional to a proper motion velocity in the scale shown in the lower-left corner. 
    }
    \label{fig:JVN_proper}
\end{figure*}
\begin{figure*}
	\includegraphics[width=\hsize]{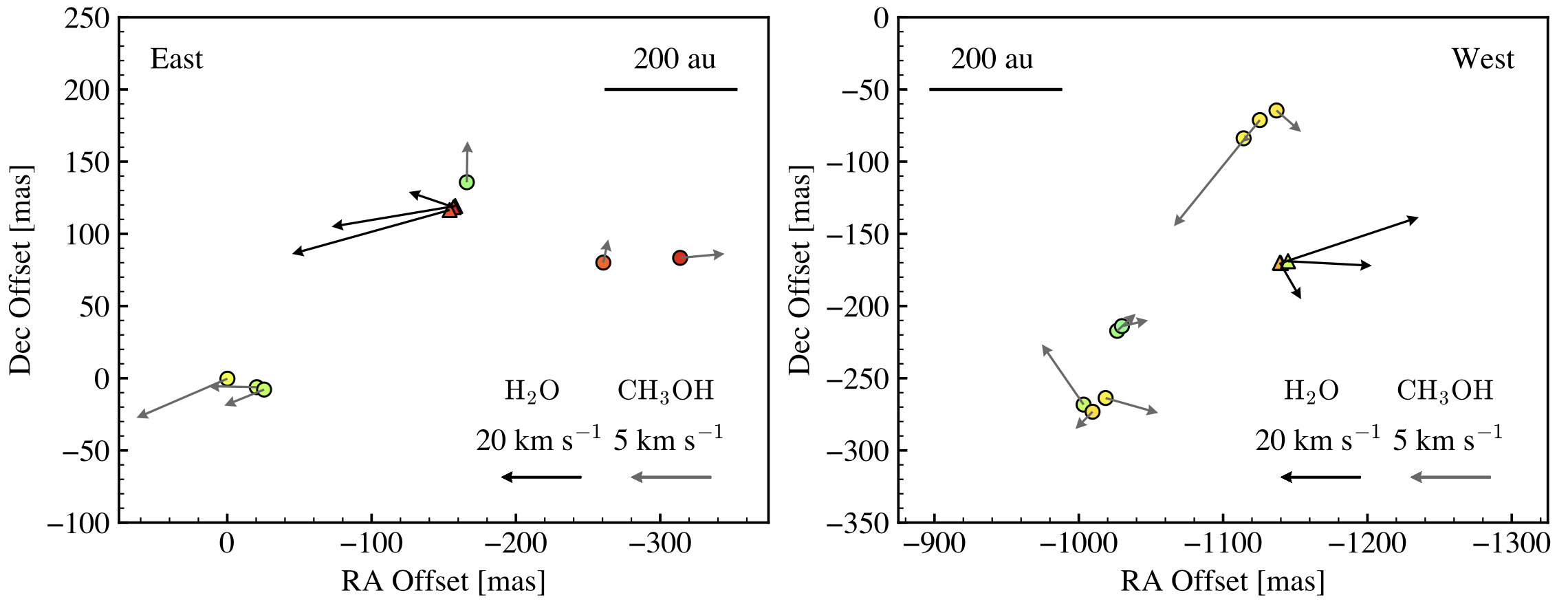}
    \caption{Close-up views of Figure \ref{fig:JVN_proper}. The left and right panels show the Eastern and Western clusters, respectively. The markers, colours, and vectors are same as Figure \ref{fig:JVN_proper}. }
    \label{fig:JVN_proper_EW}
\end{figure*}

We have detected 102, 60, 63, and 4 H$_2$O maser spots in the first to fourth epoch, respectively. 
Figure \ref{fig:VERA_map_fig} shows the spatial distributions of the detected H$_2$O maser spots in each epoch. 

\begin{table*}
    \centering
    \caption{Comparison with the previously published maser data}
    \label{tab:compare_evn_mer}
    \begin{tabular}{lcccccc}  
        \hline\noalign{\vskip3pt} 
        Type  & CH$_3$OH & CH$_3$OH & CH$_3$OH & H$_2$O &  H$_2$O &  H$_2$O \\ \hline
        Array & MERLIN & EVN & JVN & MERLIN & VERA & e-MERLIN \\   [2pt] 
        Observation date  & 2004 Dec. 31  & 2010 Mar. 14   & 2016 Aug. 25  &  2005 Jan. 1 & 2016. Feb. 10 & 2019 Mar. 20\\  
        Synthesised Beam (mas$^2$) &
        70$\times$45   & 6.5$\times$3.8   & 4.76$\times$3.09  & 20$\times$20 & 1.25$\times$0.75 & 29$\times$28 \\
        Channel spacing (km s$^{-1}$)  & 0.04   & 0.089 & 0.176 & 0.21 & 0.21 & 0.105  \\
        1$\sigma$ RMS (Jy beam$^{-1}$)    & 0.0346     & 0.006    & 0.02 & 0.0707 & 0.122 & 0.050 \\
        Numbers of features (East:West)$^{\rm a}$ & 20 (14:6) & 24 (10:14)   & 14 (6:8) & 5 (0:5) &  6 (3:3) & 4 (0:4)   \\
        References$^{\rm b}$ & 1 & 2 & This work & 1 & This work & 1 \\ 
        \hline\noalign{\vskip3pt} 
        \multicolumn{7}{l}{$^{\rm a}$ Two parenthesis digits separated by comma show numbers of maser features in the Eastern and Western cluster, respectively.}\\
        \multicolumn{7}{l}{$^{\rm b}$ 1: \citep[][]{Darwish2020}, 2: \citep[][]{Bart2014}.}\\
    \end{tabular}
\end{table*}

\begin{table*}
    \centering
    \caption{Absolute positions of referenced maser spots measured by different interferometers}
	\label{tab:absolute_position}
	\scalebox{1.0}{
    \begin{tabular}{llccll}  
        \hline\noalign{\vskip3pt} 
        & & &$V_{\rm LSR}$$^{\rm a}$ & \multicolumn{2}{c}{RADEC (J2000.0)}  \\  
        Species & Array & Year& (km s$^{-1}$) & ($\mathrm{19^{h} 43^{m}}$) & ($\mathrm{+23^{\degr}} 44'$) \\ \hline
    \multicolumn{6}{c}{Brightest H$_2$O maser features} \\ \noalign{\vskip1.5pt} 
    $\mathrm{H_2O}$ & MERLIN  & 2005 &  26.42  & $\mathrm{11^{s}.1812}$ & $\mathrm{02''.890}$  \\ 
                         & VERA & 2016 &  17.36 & $\mathrm{11^{s}.1876}$ & $\mathrm{03''.009}$ \\
                         & e-MERLIN & 2019 &  23.39 & $\mathrm{11^{s}.1844}$ & $\mathrm{03''.008}$ \\ \noalign{\vskip1.5pt} 
    \multicolumn{6}{c}{Identical CH$_3$OH maser features} \\ \noalign{\vskip1.5pt} 
    $\mathrm{CH_3OH}$   & MERLIN & 2004 &  18.97 & $\mathrm{11^{s}.2688}$ & $\mathrm{03''.218}$  \\ 
                         & EVN  & 2010  &  19.13 & $\mathrm{11^{s}.270}$ &$\mathrm{03''.2053}$ \\
                         & JVN  & 2016 & 19.25 & $\mathrm{11^{s}.2710}$ & $\mathrm{03''.178}$ \\ 
    \hline 
    \multicolumn{6}{l}{$^{\rm a}$ $V_{\rm LSR}$ of the peak maser spots.}\\
    \end{tabular}
    }
\end{table*}

We found two maser clusters in the East and West, forming `pair'-like distribution in the initial three epochs. 
This pair-like distribution differed from the previous MERLIN data, which detected only the Western cluster \citep{Darwish2020}. 
In the fourth epoch, we observed a similar monopolar distribution as the Eastern cluster disappeared. 
We note that the Western maser spots also decreased in this epoch, leaving only a single maser feature detected. 
Since the phase-referenced H$_2$O maser feature also disappeared, we estimated the relative position of the detected maser spots in the fourth epoch by extrapolating a measured proper motion (see below). 
Figure \ref{fig:VERA_map_fig_zoom1} and \ref{fig:VERA_map_fig_zoom2} shows a zoomed-up view of each maser cluster. 
Individual clusters show a mostly linear distribution along the SE-NW direction, which can be interpreted in terms of shock fronts caused by the NE-SW outflow, as seen on larger scales.

We have detected 72, 90, 62, 134, 149, and 157  CH$_3$OH maser spots in the first to sixth epoch, respectively. 
Figure \ref{fig:JVN_map_fig} presents the spatial distributions of the detected CH$_3$OH maser spots in each epochs. 
The overall distribution remained stable throughout all the observing epochs, although some individual maser spots were short-lived. 
We found the E-W `pair' distribution similar to the H$_2$O masers. This pair distribution is consistent with the previous MERLIN and EVN data \citep[][]{Darwish2020,Bart2014}. 

All the identified H$_2$O and CH$_3$OH maser features are listed in Table \ref{Feature_VERA_JVN}, with the measured parameters at the first epoch. 
The complete lists of parameters for all features are available in appendix \ref{all_feature_table} (see Table \ref{tab:all_feature_table_w} and \ref{tab:all_feature_table_m}). 
We assumed that the single H$_2$O maser feature in the fourth epoch was identical with w-4 feature, because of the similar LSR velocity. 
Three Eastern H$_2$O features (w-1, 2, and 3) are closely located to a single CH$_3$OH feature (m-4) within $\sim$ 20 mas, although they may be at different distances along the LoS since the physical conditions required for each maser species are different.

\subsection{Internal proper motions}
6 H$_2$O and 14 CH$_3$OH features were available for proper motion measurements. 
We estimated internal proper motions relative to the `barycentre' \citep[e.g.,][]{Sanna2010A&A...517A..78S, Sugi2016} since the absolute coordinate was obtained only in the first epoch of each maser species. 
We defined the barycentre of each maser species by using long-lived maser features which were detected in all the epochs, except for the fourth VERA epoch where only a single H$_2$O maser feature was available.  
In each epoch, we first calculated two averaged positions, one for the Eastern maser cluster and another for the Western maser cluster, based on the detected maser features in each cluster. 
We then obtained the overall averaged position of the two clusters, avoiding any bias from the number of detected features in each cluster. 
This overall averaged position is referred to as the barycentre.

\begin{figure*}
	\includegraphics[width=\hsize]{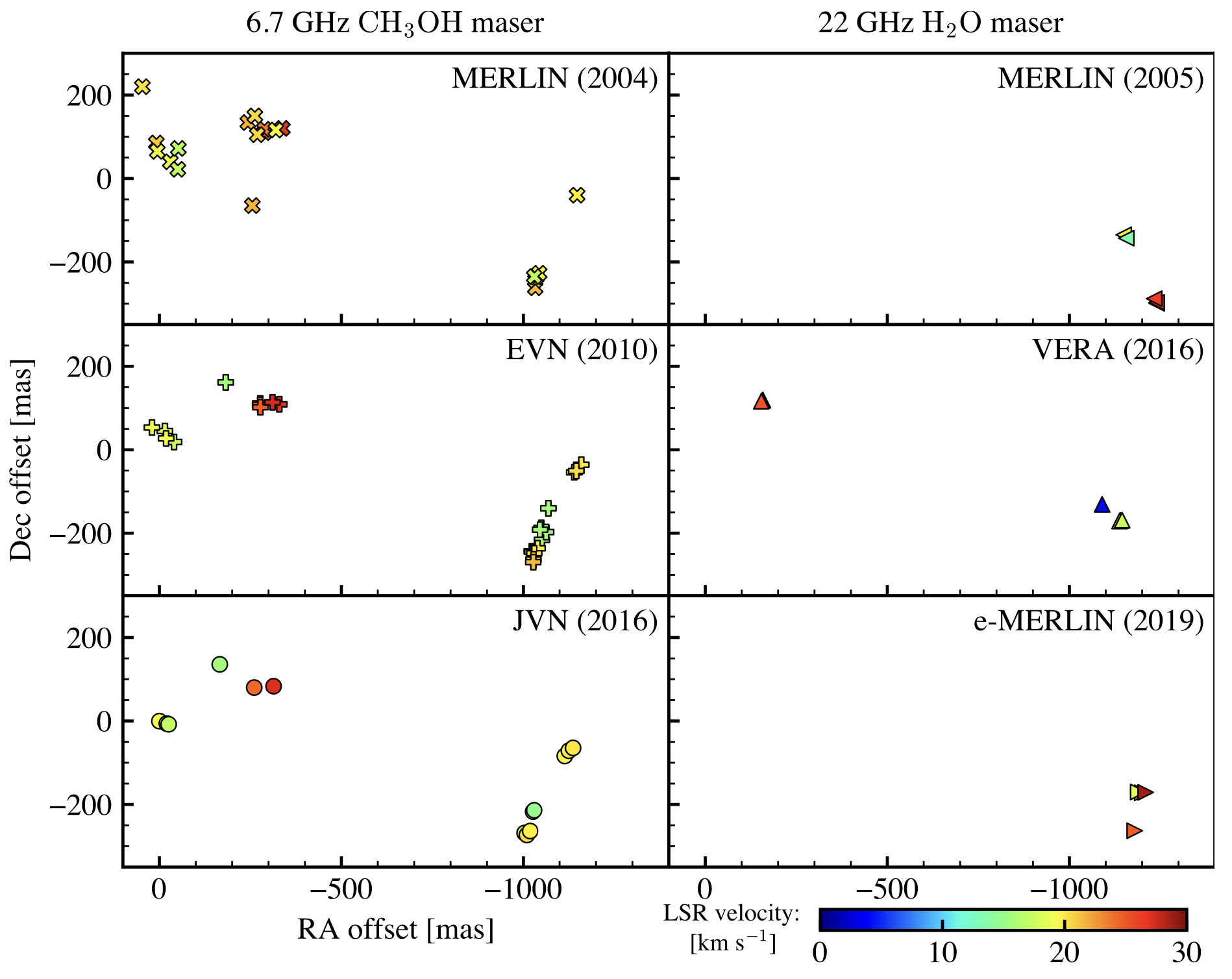}
    \caption{
    Spatial distribution of previously reported CH$_3$OH and H$_2$O maser features. 
    Different symbols represent different arrays used for the observations. 
    The colour of each marker represents the LSR velocities. The coordinate origin is the same as in Figure \ref{fig:JVN_proper}. We showed the locations of both long- and short-lived maser features at the first epoch of our data}. 
    \label{fig:map_each_fig}
\end{figure*}

Table \ref{barycenter_H2O} and \ref{barycenter_CH3OH} list the barycentres in each epoch for the H$_2$O and CH$_3$OH masers, respectively. 
We note that the barycentres of H$_2$O masers were estimated for the first three epochs,  as only the Western cluster was detected in the fourth epoch.  
Linear proper motions of each maser feature in RA ($\mu_\alpha$) and DEC ($\mu_\delta$) were estimated by the least-squares fitting, by aligning the barycentres measured at different epochs for each maser species separately. 
The estimated proper motions ($\mu_\alpha$, $\mu_\delta$), corresponding linear velocities ($V_{\rm RA}$, $V_{\rm DEC}$), total proper motion velocities ($V_{\rm prop}$), and internal LoS velocities ($V_{\rm LoS}$) are listed in Table \ref{proper_motion_VERA_JVN}. 
$V_{\rm LoS}$ indicates the offsets of the LoS velocities from the systemic velocity (22.4 km s$^{-1}$). 
A proper motion of 1 mas yr$^{-1}$ corresponds to a linear velocity of 10.24 km s$^{-1}$ at 2.16 kpc. 
We extrapolated the position of w-4 feature at the fourth epoch by using the proper motion in Table \ref{proper_motion_VERA_JVN} as mentioned in section 3.1. 

Figure \ref{fig:JVN_proper} presents the estimated proper motions of each maser feature. 
We also showed a close-up view of the Eastern and Western clusters in Figure \ref{fig:JVN_proper_EW}. 
The proper motions of H$_2$O masers range from 10.5 to 40.8 km s$^{-1}$ with an average of 25.3 km s$^{-1}$.  
All six H$_2$O features exhibit an expanding motion away from the barycentre in a NE-SW bipolar manner. 
This expanding motion indicates that the H$_2$O masers in G59 trace the root of the bipolar outflow along the NE-SW direction, which was previously observed on a larger scale. 

The proper motions of CH$_3$OH masers range 0.4 to 8.5 km s$^{-1}$, with an average of 3.0 km s$^{-1}$. 
This velocity range is significantly slower than that of the H$_2$O masers, despite both maser species exhibiting a similar bipolar distribution. 

\begin{table*}
\caption{Identical maser features between JVN, EVN, and MERLIN data}
\label{tab:offset}
\begin{tabular}{llcccccccccrrrr}
\hline
& \multicolumn{4}{c}{} & \multicolumn{3}{c}{} & \multicolumn{4}{c}{} & \multicolumn{2}{c}{Positional offsets} \\
Name & \multicolumn{4}{c}{JVN}             & \multicolumn{3}{c}{EVN}      & \multicolumn{3}{c}{MERLIN(MLN)} & \multicolumn{2}{c}{EVN - JVN } &  \multicolumn{2}{c}{MLN- JVN} \\\hline
     & $V_\mathrm{LSR}$  & $\Delta V$  & $\Delta$RA & $\Delta$DEC  & $V_\mathrm{LSR}$  & $\Delta$RA & $\Delta$DEC & $V_\mathrm{LSR}$  & $\Delta$RA & $\Delta$DEC  & RA & DEC   & RA          & DEC         \\
 & \multicolumn{2}{c}{(km s$^{-1}$)} & (mas) & (mas) & (km s$^{-1}$) & (mas) & (mas) & (km s$^{-1}$) & (mas) & (mas) & \multicolumn{2}{c}{(mas)} & \multicolumn{2}{c}{(mas)}\\ \hline
m-1  & 19.25 & 1.06 & -0.121    & -0.331   & 19.13 & -19.003   & 27.284   & 18.97 & -30   & 40   & -19 & 28 & -30 & 40  \\
m-2  & 17.32 & 0.36 & -20.445   & -6.197   & 17.20 & -40.833   & 18.673   & 17.08 & -51   & 22   & -20 & 25 & -30 & 28  \\
m-3$^{\rm a}$  & 17.06 & 0.53 & -25.529   & -7.844   &  &  &  &  &  &  & (-15) & (27) & (-25) & (30) \\
m-4  & 15.56 & 0.35 & -165.977  & 135.722  & 15.70 & -182.776  & 161.502  & -     & -     & -    & -17 & 26 &  -  & -    \\
m-5  & 24.70 & 1.23 & -260.561  & 80.099   & 24.75 & -278.049  & 101.954  & 24.46 & -288  & 110  & -17 & 22 &  -28 & 30  \\
m-6  & 27.15 & 1.05 & -313.786  & 83.363   & 27.03 & -330.000  & 109.000  & 26.92 & -338  & 120  & -16 & 26 &  -24 & 37  \\
m-7  & 17.06 & 0.53 & -1003.316 & -268.261 & 17.20 & -1025.021 & -244.759 & 16.95 & -1030 & -235 & -22 & 24 &  -27 & 33  \\
m-8  & 20.22 & 0.53 & -1009.536 & -273.207 & 20.71 & -1028.311 & -262.858 & 20.60 & -1033 & -244 & -19 & 10 &  -23 & 29  \\
m-9  & 19.78 & 0.35 & -1018.565 & -263.700 & 19.83 & -1039.908 & -236.120 & 19.67 & -1044 & -227 & -21 & 28 &  -25 & 37  \\
m-10 & 15.48 & 0.87 & -1026.591 & -217.185 & 15.53 & -1046.508 & -191.663 & -     & -     & -    & -20 & 26 &  -  & -    \\
m-11 & 14.69 & 0.35 & -1029.893 & -213.909 & 14.56 & -1050.339 & -187.368 & -     & -     & -    & -20 & 27 &  -  & -    \\
m-12 & 19.52 & 0.53 & -1114.213 & -83.831  & 19.74 & -1140.734 & -54.106  & 19.72 & -1148 & -40  & -27 & 30 &  -34 & 44  \\
m-13 & 19.78 & 0.70 & -1125.450 & -71.219  & 20.18 & -1146.398 & -50.881  & -     & -     & -    & -21 & 20 &  -  & -    \\
m-14 & 19.95 & 0.70 & -1136.994 & -64.560  & 20.01 & -1160.068 & -35.758  & -     & -     & -    & -23 & 29 &  -  & -    \\\hline  
\multicolumn{8}{l}{$^{\rm a}$ m-2 and m-3 could be identical with the same EVN/MERLIN features (see main text).} &\multicolumn{3}{l}{(i): Averaged offset } & -20 & 24  & -28 & 35
\\\noalign{\vskip3pt} \cline{9-15} \noalign{\vskip3pt} 
\multicolumn{8}{l}{}&\multicolumn{3}{l}{(ii): Expected secular motion}  & 11 & 33  & 20 & 59
\\\noalign{\vskip3pt} \cline{9-15} \noalign{\vskip3pt} 
\multicolumn{8}{r}{}&\multicolumn{3}{l}{(i) - (ii): Residuals} & -31 & -11 & -48 & -24 
\\ \cline{9-15} 
\end{tabular}
\end{table*}

\section{Discussions}

\begin{figure*}
	\includegraphics[width=0.8\hsize]{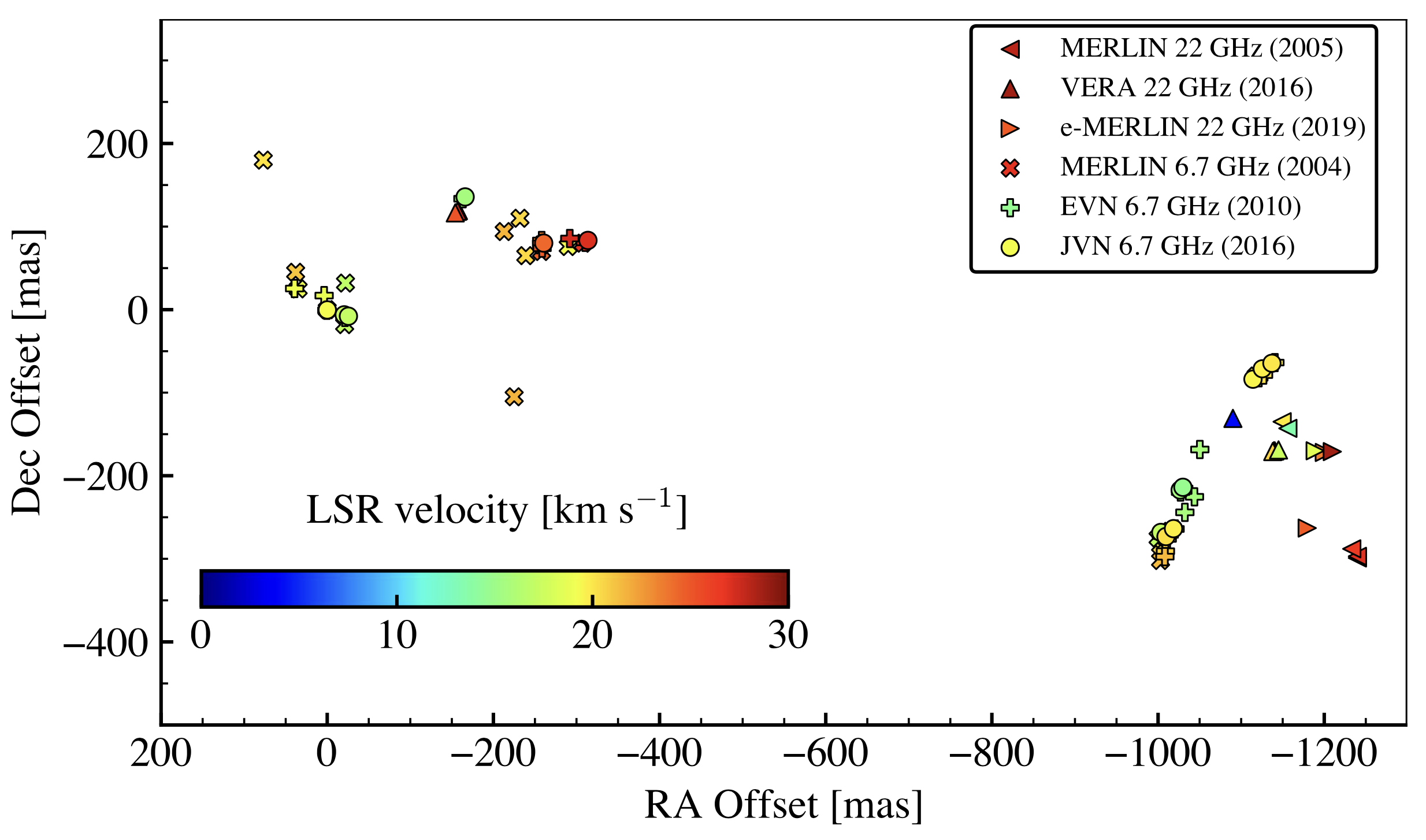}
    \caption{
    All the maser data are plotted simultaneously in relative coordinates with respect to the maser feature m-1. 
    }
    \label{fig:multi_default_fig}
\end{figure*}

\begin{figure*}
	\includegraphics[width=\hsize]{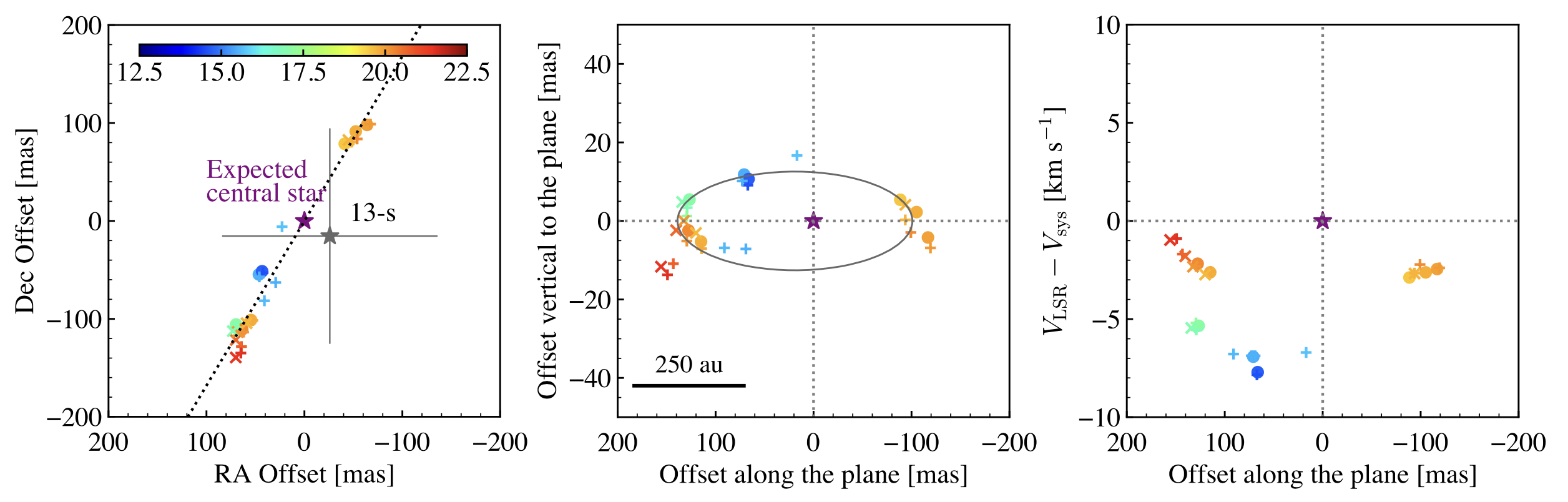}
    \caption{Left: close-up view of the Western CH$_3$OH masers in Figure \ref{fig:multi_default_fig}. 
    We used the same markers for the CH$_3$OH masers, but the velocity range shown by the colour bar is slightly narrower than Figure \ref{fig:multi_default_fig}. 
    The black dotted line represents the result of linear fitting on the CH$_3$OH masers. 
    The coordinate origin is the expected position of the host protostar, marked by the violet star. 
    This position is the nearest point on the fitted line from the dust continuum peak at 1.4 mm, named `13-s' (see main text). 
    The grey star and cross indicate the original position and positional error of 13-s in \citet{Rod2012}, respectively. 
    Middle: Further zoomed-up view of the `linear' distribution. 
    The horizontal axis is along the fitted line centred on the stellar position, increasing towards the South-East side.
    The vertical axis shows height from the line, increasing towards the North-East side, which is four times magnified than the horizontal axis. 
    The black ellipse was fitted by the inner ring-like masers. 
    Right: $PV$-diagram along the fitted line. The horizontal axis shows internal LoS velocities ($V_\mathrm{LSR} - V_\mathrm{sys}$). 
    }
\label{fig:pv_obs_fig}
\end{figure*}

\subsection{Comparison with the previous (e-)MERLIN and EVN data}
Table \ref{tab:compare_evn_mer} summarises observing parameters from available interferometric mapping data over the last two decades. 
Table \ref{tab:absolute_position} lists the absolute coordinates of referenced maser spots with different interferometers at different times. The alignment of maser positions among different observations will be commented on further below. 
Figure 7 presents the spatial distributions of each mapping data with offsets relative to the position of the CH$_3$OH reference (same as Figure \ref{fig:JVN_proper}). 

We found a very similar spatial distribution of CH$_{3}$OH masers. This fact allows us to identify 13 long-lived maser features within the velocity width of the JVN data. 
We first associated maser features with consistent LSR velocities within the velocity width of each feature. We then confirmed that the relative feature positions in each dataset were consistent. 
Table \ref{tab:offset} shows the maser features also detected by EVN and/or MERLIN. 
We note that both the m-2 and m-3 features in the JVN data could be associated with the same EVN/MERLIN features. Although we chose the m-2 feature in this paper, they were closely located, and this choice did not significantly affect our estimation of averaged proper motion. 
We observed systematic offsets in absolute coordinates between the JVN, EVN and MERLIN data. The averaged offsets of the EVN and MERLIN data are also listed in Table \ref{tab:offset}. 
\citet{xu2009} measured the secular motion of G59 by using the CH$_3$OH masers at 12 GHz, i.e., -1.7 and -5.1 mas yr$^{-1}$ in RA and DEC, respectively. 

Their secular motion corresponds to expected positional shifts of (+11 mas, +33 mas) for EVN and (+20 mas, +59 mas) for MERLN, when comparing with the JVN positions in RADEC. 
The average offsets in Table \ref{tab:offset} are inconsistent with these expected values. 
We, thus, consider that there is a significant astrometric error in the absolute coordinate measured among the different epochs (up to 50 mas). 
Hereafter, we compared these three data in a relative coordinate with respect to positions of the maser feature m-1 in Table \ref{tab:offset}. 
 
Figure \ref{fig:multi_default_fig} presents all the maser data in a single plot. 
Overall distributions of CH$_3$OH have not changed over 15 years from 2004 to 2019. 
This fact implies that the physical environment for pumping and amplification of these CH$_3$OH maser has remained stable in G59. 
The H$_2$O masers were, on the other hand, basically concentrated in the Western cluster. 
The bipolar `pair' distribution was only detected in the first to third epochs of our VERA data (see Figure \ref{fig:VERA_map_fig} again). 
Such variability is a common feature of H$_2$O masers excited in protostellar outflow and, more generally, in shocks, which trace highly episodic mass ejection in some cases \citep[e.g.,][]{Motogi2011MNRAS,Motogi2016PASJ,Burns2016}.

\subsection{Position-Velocity diagram for the Western masers}
We have found that the Western CH$_3$OH masers exhibit a `linear'-like distribution along the SE-NW direction (see the left panel in Figure \ref{fig:pv_obs_fig}). 
The millimetre continuum peak `13-s' reported in \citet{Rod2012} is nearly located at the centre of this distribution.  
We have inspected whether this linear distribution might possibly trace the edge-on disc-like structure or not, as follows. 

We first made a linear fitting on this maser distribution. 
The PA of the best-fit line was determined to be 149$\degr$ from North to East. 
We then assumed that the host protostar is located at the nearest position from 13-s on the best-fit line. 
This position was set to the origin in Figure \ref{fig:pv_obs_fig}. 
The middle panel of Figure \ref{fig:pv_obs_fig} presents a further zoomed-up view of the Western CH$_3$OH masers, which has been rotated so that the horizontal axis coincides with the best-fit line. 
The vertical axis is perpendicular to the line and is four times magnified compared to the horizontal axis. 
This scale allows us to highlight deviations from a linear distribution in Figure \ref{fig:pv_obs_fig}, and shows that the CH3OH masers could rather trace a highly-elongated ellipse. 
We also showed the best-fit ellipse in Figure \ref{fig:pv_obs_fig} as a reference. 
This ellipse is estimated by using the inner masers around the origin. 
The aspect ratio of the ellipse is about 12 (120 and 10 mas on the major and minor axis, respectively). 
If the ellipse traces the outer radius of the inclined disc-like structure, this aspect ratio corresponds to inclination angle of $\sim$6$\degr$ (0$\degr$ for a completely edge-on case). 
It should be noted that this inclination is small and affects only 1 per cent of LoS velocities. 
We, therefore, ignored it in the following analysis.

\begin{figure*}
\includegraphics*[width=\hsize]{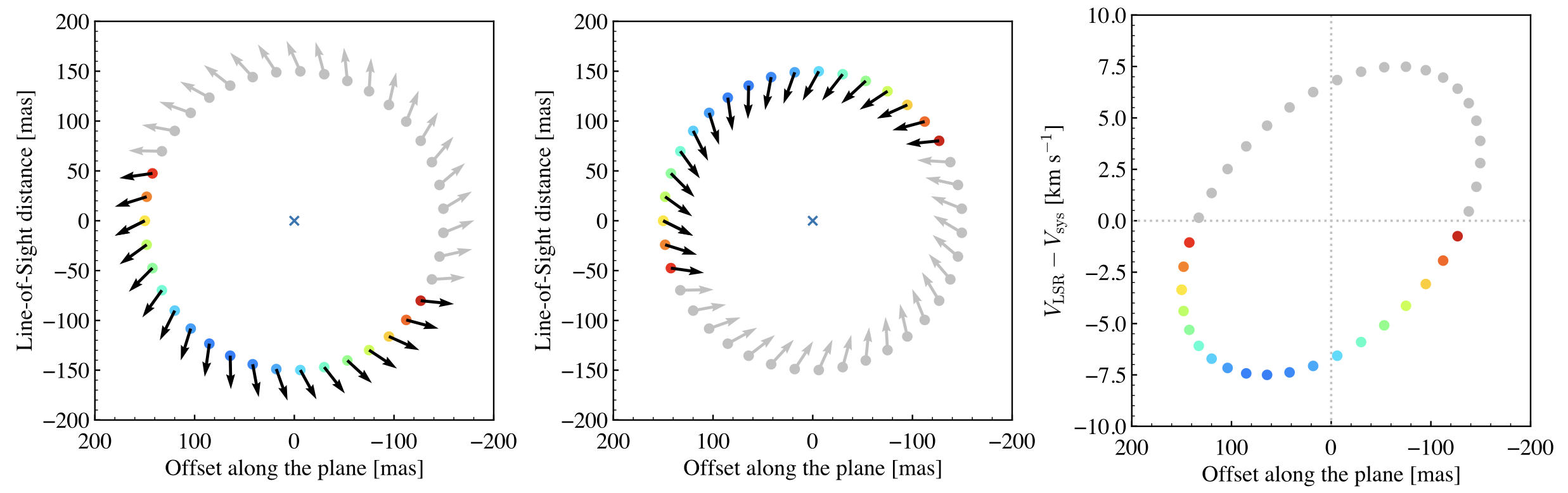}
\caption{
Face-on view of two possible kinematic models that match our $PV$ diagram. 
The left panel presents the expanding-rotating ring model, while the middle panel shows the infalling-rotating ring model. The right panel displays the $PV$ diagram when these ring models are observed edge-on. 
Both kinematic models can reproduce identical $PV$ diagrams. 
}
\label{fig:pv_model_fig}
\end{figure*}

The right panel in Figure \ref{fig:pv_obs_fig} presents the position-velocity ($PV$) diagram along the best-fit line. 
The horizontal axis shows the positional offset from the origin, increasing towards the SE side, while the vertical axis shows internal LoS velocities. 
The CH$_3$OH masers exhibit a half ellipse only on the blue-shifted side in the $PV$ diagram. 
This U-shaped $PV$ diagram typically appears as a part of a rotating expanding or rotating infalling ring \citep[e.g., ][]{Hirota2017NatAs}. 
In a previous analysis of the $PV$ diagram of the Western CH$_3$OH masers in G59, \citet{Phillips2000evn} focused only on the Southern side of our best-fit line. They ignored the Northern side since only a single maser feature was detected there. They investigated Keplerian rotation, considering a linear velocity gradient on the Southern side, which is indeed observed in the third quadrant of our $PV$ diagram.

\subsection{Rotating-infalling/expanding ring model}
We conducted a kinematic fitting on the $PV$ diagram, assuming a rotating-infalling/expanding ring model (see Appendix \ref{ellipse_fit}).  
In this model, we assumed that the host protostar itself has no additional LoS motion with respect to the natal cloud (i.e., $V_{\rm sys}$ of 22.4 km s$^{-1}$). 
We found that if we assumed an edge-on ring, either rotation with infall or rotation with expansion could reproduce the observed $PV$ diagram (see Figure \ref{fig:pv_model_fig}). 
Both models are rotating counterclockwise, but the location of CH$_3$OH masers along the LoS differs from each other. 
In the infalling model, the masers are located on the far side of the ring, while in the expanding model, they are situated on the near side. 

\begin{figure}
    \centering
	\includegraphics[width=\hsize]{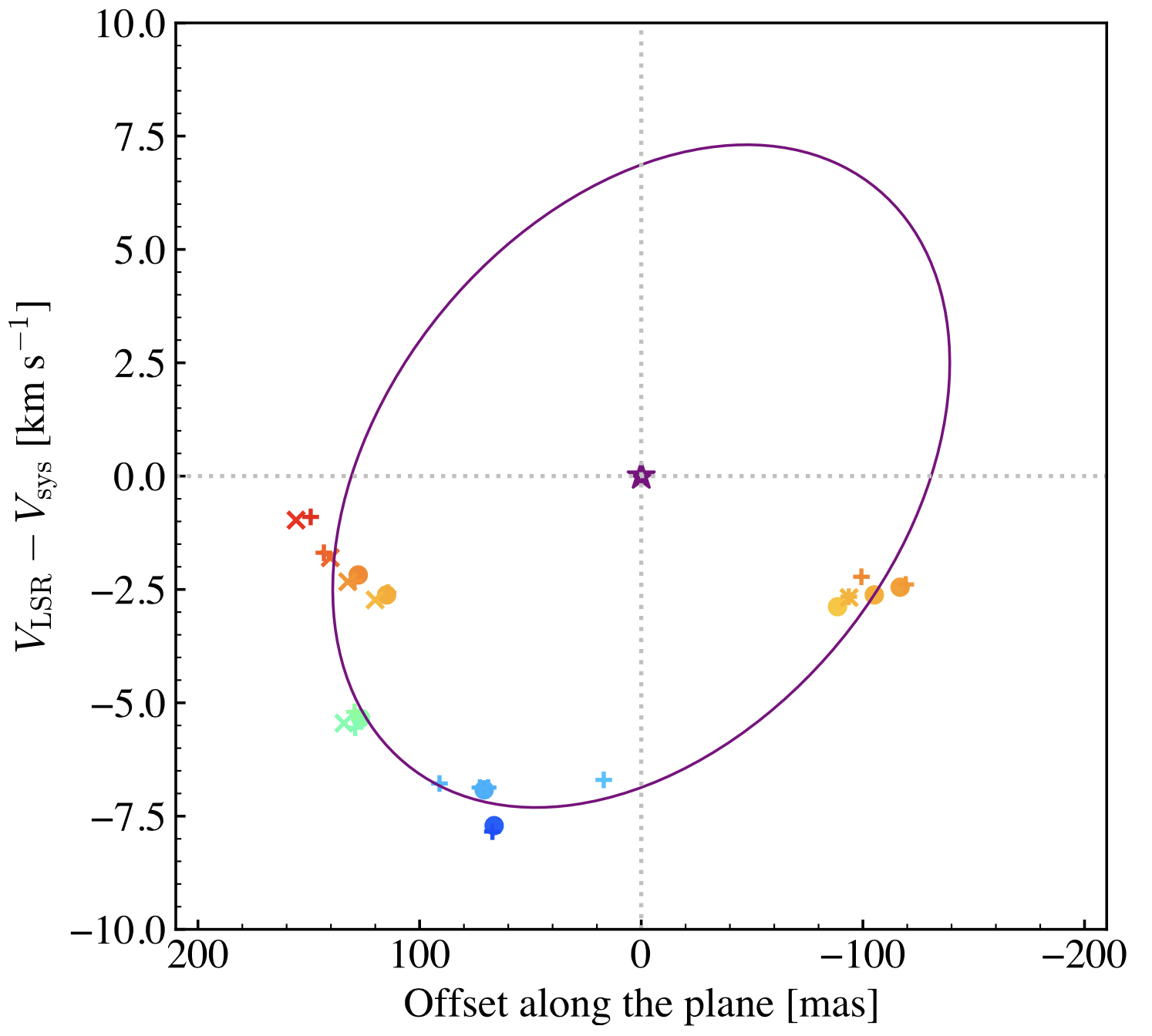}
    \caption{The best-fit $PV$ model for the Western CH$_3$OH masers. The derived parameters are summarised in Table \ref{tab:pv_parameter}}
    \label{fig:LS_fit_fig}
\end{figure}

\begin{figure}
	\includegraphics[width=\hsize]{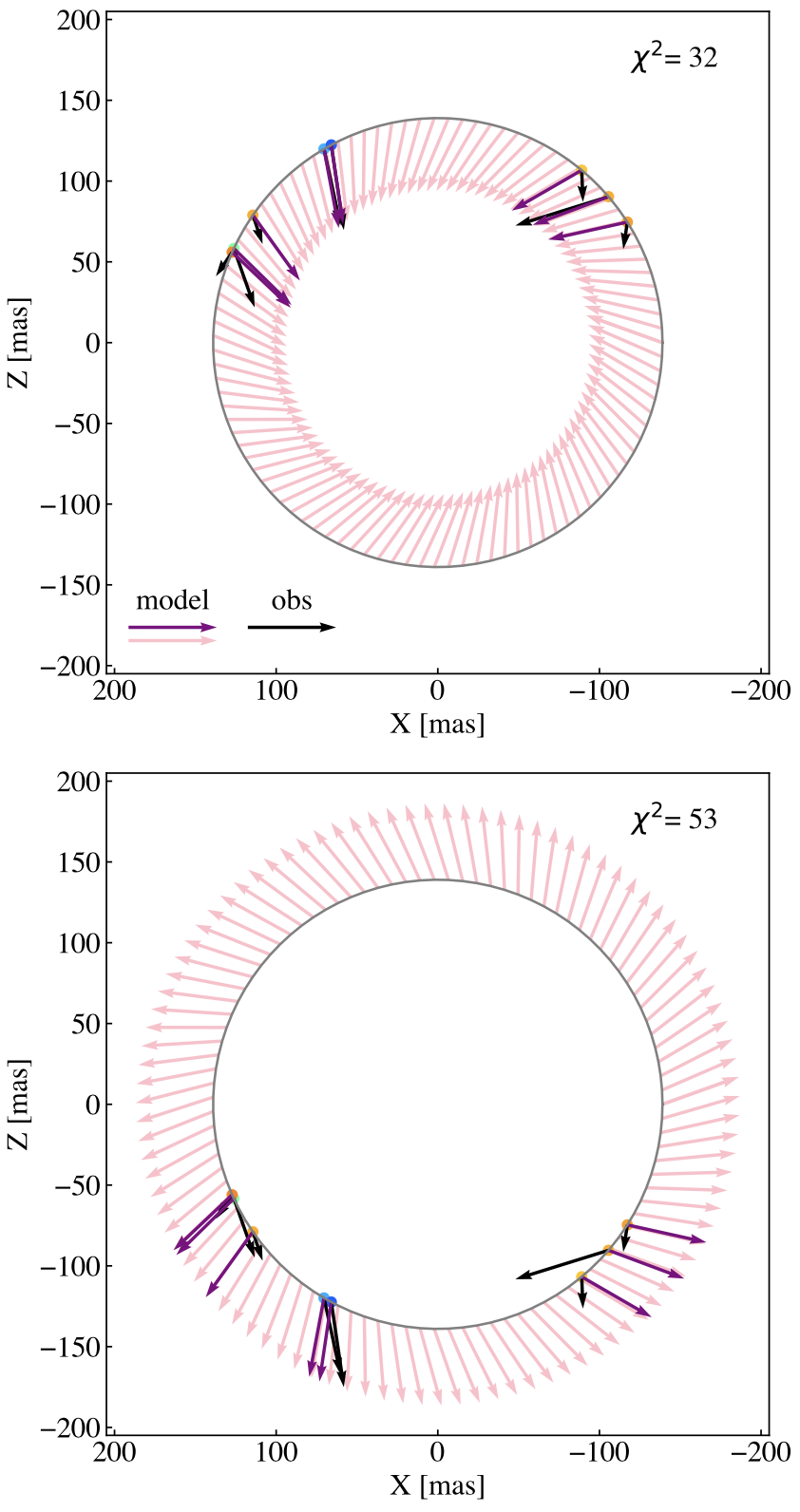}
    \caption{Comparison between the observed (black arrow) and modelled (pink and purple arrow) motion vectors in a face-on view. The top and bottom panels show the rotating infalling and expanding models, respectively, along with the corresponding $\chi$-squared values. 
    The colour-filled circles show the positions of observed maser features detected by JVN. The grey circle presents the best-fit ring of radius $r$ in Table \ref{tab:pv_parameter}. The horizontal axis ($X$) shows the positional offset along the plane of the edge-on ring. The vertical axis ($Z$) shows the position along the LoS, which is given by $Z = \sqrt{r^2-X^2}$. 
    The purple arrows highlight the expected motions of each maser feature, with the motion vectors projected on the $X$ axis representing the proper motions along the line of the ring. 
    }
    \label{fig:ER_IR_fig}
\end{figure}

Figure \ref{fig:LS_fit_fig} shows the best-fit ellipse model on the $PV$ diagram. 
Table \ref{tab:pv_parameter} lists the best-fit parameters. 
The maser ring has a radius of 139 $\pm$ 5 mas (300 $\pm$ 11 au at 2.16 kpc), a rotating velocity of 2.5 $\pm$ 0.3 km s$^{-1}$, and expanding or infalling velocity of 6.9 $\pm$ 0.5 km s$^{-1}$. 
This result, at least, indicates that the maser kinematics is not consistent with a scenario of a simple rotationally-supported disc. 
Since we could not favour one model over the other only by the $PV$ diagram, we next inspected proper motion vectors. 
Figure \ref{fig:ER_IR_fig} compares the observed and modelled (proper) motion vectors. 
For this comparison, we have recalculated observed proper motions relative to the centre of the Western maser clusters that is almost consistent with the expected position of the host protostar. 
We only compared the proper motion vectors along the plane of the ring since our model did not include any motions perpendicular to this plane. 

We calculated $\chi^2$ values as follows: 
\begin{equation}
\chi^2 = \sum_{i} \frac{(\mu_{Xi}^{\rm obs} - \mu_{Xi}^\mathrm{mod})^2}{\sigma_i^2},
\label{eq:chi2}
\end{equation}
where $\mu_{Xi}^{\rm obs}$ and $\mu_{Xi}^{\rm model}$ are the observed and modelled proper motions along the plane of the ring for the i-th maser feature, respectively.
$\sigma_i$ indicates the error in the observed proper motion for the i-th maser feature. 
We found that the infalling model gives better fitting than the expanding model, giving a factor of $\sim$ 2 better $\chi^2$. 

These fitting results may not be conclusive but they are consistent with \citet{Darwish2020}, who suggested that the Western CH$_3$OH masers were excited in infalling gas. 
In this case, the CH$_3$OH masers are located outside the centrifugal barrier or selectively tracing infalling components caused by angular momentum transfer in the inner circumstellar regions \citep[e.g.,][]{Motogi2017ApJ,Motogi2019}. 
A complementary $PV$ analysis of thermal emission lines with high excitation energies, that would trace the inner circumstellar regions more uniformly, is needed to confirm the presence of infall. 

On the other hand, a rotating-expanding motion of the CH$_3$OH maser has been reported in several cases \citep[e.g., ][]{Sugi2016,Bart2020}. 
The origin of such a rotating-expanding motion is still an open question. 
In our case, some CH$_3$OH maser features located on the outer edge show proper motions outward from the plane of the ring. 
This motion may suggest that these CH$_3$OH masers trace the root of the rotating disc wind, as suggested by \citet{Sugi2016}. 
However, a detailed outflow model and more precise proper motion measurements are required for a quantitative analysis, which is beyond the scope of this work. 

There are several CH$_3$OH masers outside the best-fit ellipse in the middle panel of Figure \ref{fig:pv_obs_fig} (e.g., masers with LoS velocities of 15.70, 17.06, and 19.95 km s$^{-1}$). 
These masers may reflect the width or thickness of the ring ($\sim$ 10 mas). 
Another possibility is that these masers trace a part of infalling and/or rotating spirals, which have been recently reported in a few CH$_3$OH maser sources \citep[e.g.,][]{Motogi2017ApJ, Chen2020, Burns2023}. 
Such spiral masers may trace the internal structures of an accretion disc or envelope. 
However, their distributions are far from disc-like ones, i.e., ring/ellipse, arc, and linear distribution. 
\citet{Motogi2017ApJ}, for example, reported CH$_3$OH masers tracing two spiral tails connecting to the inner compact ring around a face-on high mass protostar G353.273+0.641. 
\citet{Burns2023} also found that the CH$_3$OH traces spiral arms inside a gravitationally unstable Keplerian disc around a high mass protostar G358.93-0.03-MM1 during an accretion burst event. They reported that the masers were excited in a limited part of the spiral arms in each epoch, and the maser emitting region moved gradually outward along the propagation of a heat wave caused by episodic accretion. 
This finding implies that simple morphological classification is insufficient to distinguish the realistic physical structure traced by CH$_3$OH masers. 
It is especially essential to be cautious when deducing a physical structure based on VLBI data, as faint or extended maser spots tend to be undetectable. 

High-sensitivity (connected) interferometer observations of thermal emission lines, with comparable angular resolutions than VLBI, are needed to provide a comprehensive map of the entire physical structure, that can fill up the gap between maser features.

\begin{table}
    \centering
    \caption{The best-fit $PV$ model parameters}
	\label{tab:pv_parameter}
    \begin{tabular}{ccc}  
        \hline\noalign{\vskip3pt} 
        $r$  &$V_\mathrm{rot} $ & $V_\mathrm{inf}$ = $-V_\mathrm{exp}$ \\  [2pt] 
               (mas)& (km s$^{-1}$)  & (km s$^{-1}$)    \\
        \hline\noalign{\vskip3pt} 
        139 $\pm$ 5& 2.5 $\pm$ 0.3 & -6.9 $\pm$ 0.5  \\        
        \hline\noalign{\vskip3pt} 
    \end{tabular}
\end{table}

\subsection{Discrepancy between the kinematic model and observed proper motions}
There is some discrepancy between the modelled and observed proper motions, despite the well-fit $PV$ analysis. 
In the following, we speculate on three possible sources of uncertainty that can generally affect maser proper motion measurements, although the first one (the Christmas tree effect) does not affect the JVN data directly but prevents us from a combination with the previous EVN and MERLIN observations. 

\subsubsection{"Christmas-tree" effect and maser structure} 
A well-known source of error in proper motion measurements is the Christmas-tree effect \citep{Genzel1981ApJ}, where closely-located maser spots switch by blinking like a Christmas tree. 
This effect causes an apparent jump in the position of the initial fading maser spot to that of another brightening maser spot between two observing epochs. 

The overall distribution of the CH$_3$OH masers in G59 has remained over fifteen years, implying that warm and dense gas suitable for maser excitation is consistently abundant at the same locations. However, only a limited volume of such gas can satisfy the coherent condition for maser amplification in general. Several maser spots should be recurrently excited and extinguished in a local gas clump. 

We could suffer from the Christmas-tree effect since there is a long gap between the third and fourth epochs. 
Although we have carefully excluded non-linear jumps in maser positions, it is difficult to identify small linear-like jumps within a range of positional errors between the first three and the latter three epochs. 

\begin{figure}
    \centering
	\includegraphics[width=1.0\hsize]{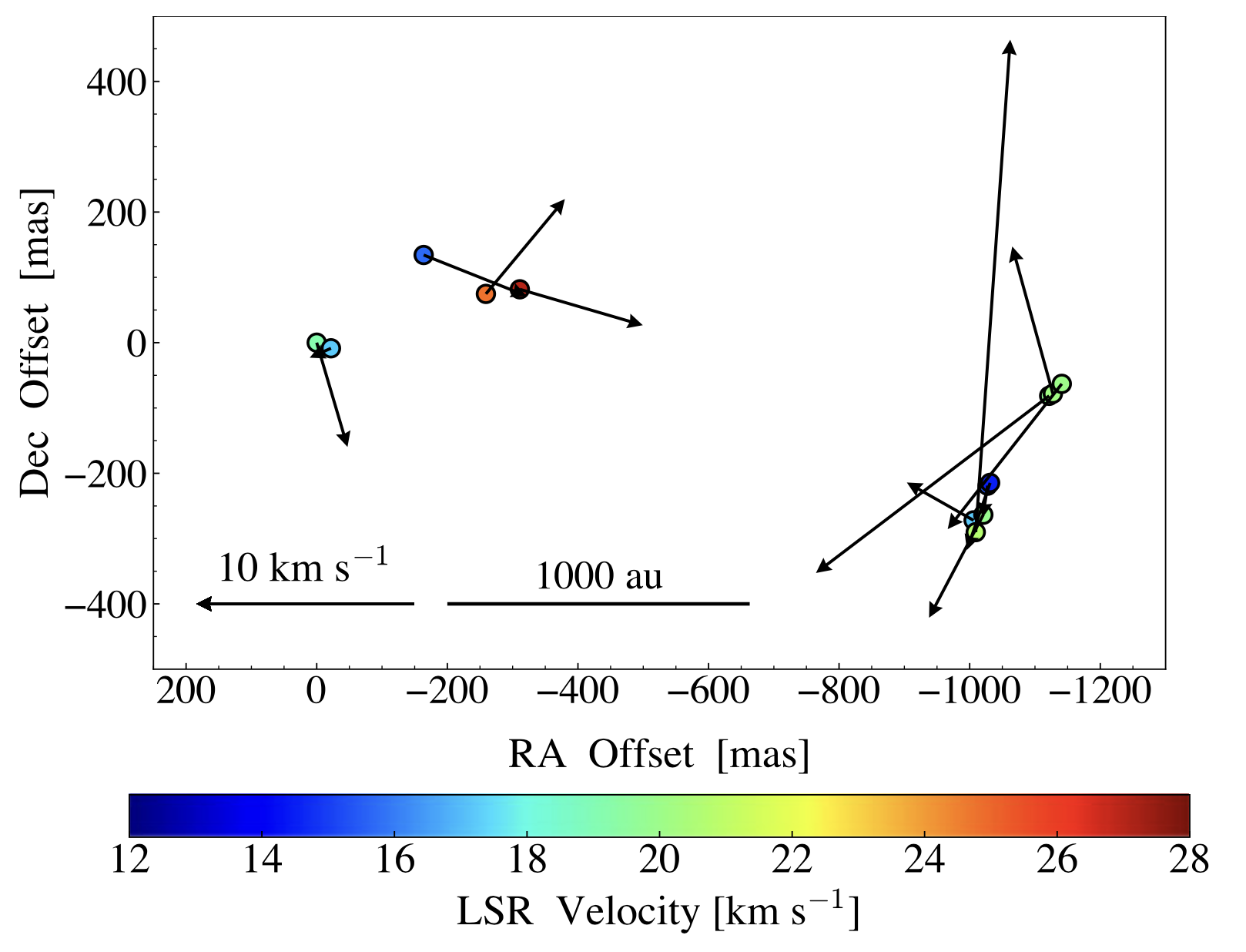}
    \caption{Long-term internal proper motions of CH$_3$OH masers in G59 derived from the EVN and JVN data. The coordinate origin is the position of the referenced maser in Figure \ref{fig:map_each_fig}. 
    } 
    \label{fig:J_E_proper}
\end{figure}

The Christmas-tree effect becomes more serious over longer durations. 
Figure \ref{fig:J_E_proper} shows the long-term internal proper motion between the JVN and EVN data.
We derived these motions for identical maser features listed in Table \ref{tab:offset} using the same procedure adopted for the JVN data. 

We recalculated the barycentres using only the identical masers. 
The new barycentres were (-624.897 mas, -37.128 mas) for EVN and (-605.124 mas, -61.726 mas) for JVN. The latter position differs slightly from that in Table \ref{barycenter_CH3OH} due to the limited number of maser features.

Prominent inward proper motions indicate that the distribution of CH$_3$OH masers shrunk between 2010 and 2016. 
However, these inward motions are significantly larger than what is expected from the $PV$ model. 
In particular, some masers showed a proper motion of $\sim$ 30 km s$^{-1}$. Such a large motion is inconsistent with our short-term proper motions and also clearly larger than the typical proper motion found for CH$_3$OH masers ($\sim$ a few km s$^{-1}$). 
These facts suggest that derived inward motions are apparent motions caused by the Christmas-tree effect. 
Hence, we avoid further discussions using the long-term proper motion combing the JVN, EVN, and MERLN data. 

The proper motions measured only by the JVN data could be more severely affected by the structural variation of individual maser features. For example, the number of maser spots varied at weak line wings in some maser features. 
However, errors due to such variations should be reduced by error-weighted averaging (see Section 2.3), as positional errors are generally larger for weaker maser spots. 
Additionally, significant changes in channel spacing (see Table \ref{JVN_obs}) can cause non-negligible positional errors for maser features with a narrow velocity range.

\subsubsection{Turbulence in disc-like structure}
Another possible error source is turbulence in the disc-like structure. 
Local turbulence could introduce randomness in proper motion vectors. 
Although turbulence inside a high-mass disc-like structure has not been well-studied, the turbulent velocity can reach transonic levels in the low-mass cases \citep[e.g.,][]{Simon2015ApJ}. 
If we assume such transonic turbulence, the typical gas temperature of the CH$_3$OH maser \citep[$\sim$ 200 K,][]{Sobolev1997AA,Cragg2005MNRAS} corresponds to a sound velocity of $\sim$ 1 km s$^{-1}$, using a mean molecular weight of 2.5. 
This velocity is 20--30 per cent of typical proper motions observed in G59 (3--5 km s$^{-1}$). 
If the turbulence is supersonic by a factor of 2, it would be enough to account for the observed randomness in proper motion vectors. 
Such turbulence is consistent with our $PV$ diagram, where observed LoS velocities have some offsets from the best-fit model, where we only assumed uniform rotating infalling/expanding motion.

\subsection{Origin of the Eastern masers}
The Eastern CH$_3$OH and H$_2$O masers exhibit a complex distribution. 
Although, there is a spatial coincidence that suggests both maser species could trace the outflow, the estimated proper motion vectors for both maser species are inconsistent. 
Some CH$_3$OH masers do show similar outflowing motions, but the H$_2$O masers generally have higher velocities compared to the CH$_3$OH masers. 
Additionally, there are a few nearby H$_2$O and CH$_3$OH maser features that are moving in opposite directions to each other. 

Such an opposite motion could be explained if CH$_3$OH masers locate across the boundary of outflow and inflow, or if they were tracing the slow side edge of a shock where the H$_2$O masers are at the tip. 
\citet{Machida2020} suggested that some part of the outflow can become gravitationally decelerated and eventually turn into inflow in such a region (i.e., `failed outflow'). 
The observed inconsistency is naturally solved if the H$_2$O masers would be excited in high-velocity gas near the outflow axis, while the CH$_3$OH masers would be excited around the outflow/inflow boundary. 
This interpretation is consistent with the previous outflow scenario proposed by \citet{Darwish2020}. 

\subsection{Radio continuum emission tracing a radio jet}
\citet{Carral1999} detected two radio continuum sources by VLA, i.e., source NE (0.7 mJy) and SW (0.3 mJy). Source NE is located near the Western maser cluster, while source SW is located ($-$5\arcsec.1, $-$2\arcsec.8) from source NE. If these sources trace an H$_\mathrm{II}$ region, we can deduce the spectral type of the host star by excitation parameter $U$, as follows \citep[e.g., ][]{Kurts1994ApJS}: 
\begin{linenomath}
\begin{align}
U \simeq 2.1 \left(\frac{D}{2.16 \ \mathrm{kpc}}\right)^{\frac{2}{3}} \left(\frac{S_\nu}{0.7 \ \mathrm{mJy}}\right)^{\frac{1}{3}} \mathrm{pc \ cm^{-2}}. 
\end{align}
\label{eq:para_u}
\end{linenomath}  
The value of $U$ for source NE corresponds a B2-type Zero Age Main Sequence (ZAMS) \citep{Panagia1973AJ}. 
However, based on a bolometric luminosity of $10^4 L_{\sun}$, a spectral type of B0.5 has been suggested for the host star \citep{Mart2008}. 
We speculate that the radio continuum emission might trace a radio thermal jet with two knots, although this scenario needs to be confirmed by future observations. 

\begin{figure*}
	\includegraphics[width=0.8\hsize]{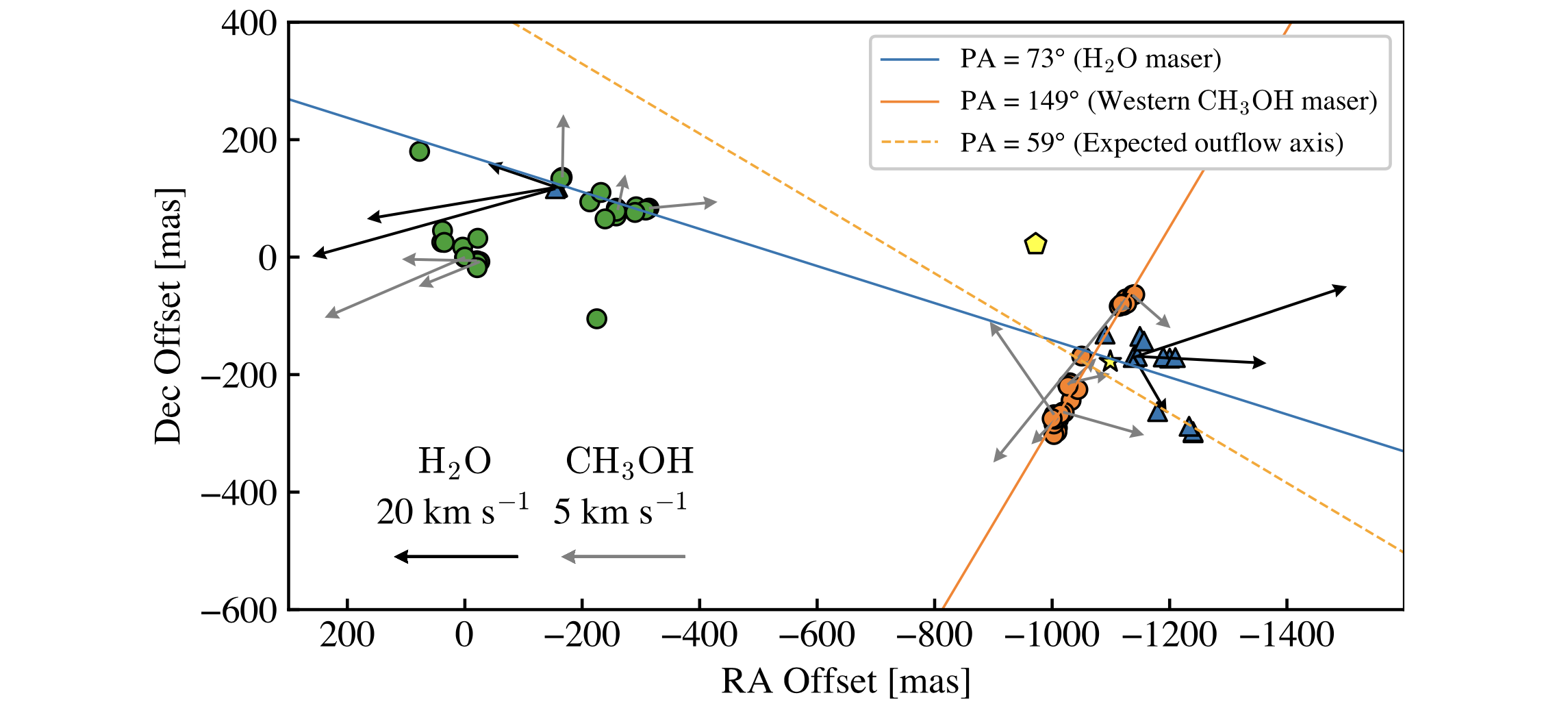}
 \caption{Comparison of the outflow axes estimated by different tracers. 
 Green circles, orange circles, and blue triangles indicate Eastern CH$_3$OH masers, Western CH$_3$OH masers, and H$_2$O masers, respectively. The star marks the peak position of the millimetre continuum 13-s. The yellow pentagon shows the position of the radio source NE. Another radio source SW is located ($-$5\arcsec.1, $-$2\arcsec.8) from NE. The black and grey arrows are the same as in Figure \ref{fig:JVN_proper_EW}. 
 The solid orange line presents PA of linearly distributed CH$_3$OH masers, interpreted as the equatorial plane of a disc-like structure. 
 The dashed orange line shows the expected outflow axis perpendicular to the linear masers, which is consistent with the PA of the possible radio jet. The solid blue line indicates the outflow direction determined by the H$_2$O masers. }
    \label{fig:G59_prop_map}
\end{figure*}

The position angle (PA) of the radio jet, calculated from the positions of the two radio sources, is 59$\degr$ in this case. 
This PA is consistent with that of the outflow cavity traced by the NIR H$_2$ line on a larger scale \citep{Mart2008}. 
Furthermore, the PA of the disc-like structure we propose, based on the maser $PV$ analysis, is 149$\degr$, i.e., just normal to this jet direction (see Figure \ref{fig:G59_prop_map}). 

These findings are consistent with a scenario where the CH$_3$OH masers and radio continuum might trace a compact disc/jet system in G59. 

Figure \ref{fig:G59_prop_map} also includes another PA of 73$\degr$, determined by a linear fitting on the distribution of H$_2$O masers. 
This direction is slightly (14$\degr$) different from the expected outflow axis. 
This discrepancy may imply that the radio jet is more directly located on the jet axis than the H$_2$O maser, similar to the case of CepA HW2 \citep[][]{Torrelles2011}. 

The positional error of source NE is, at least, better than 0$\arcsec$.2, which is given by the synthesised beam ($\sim$3$\arcsec$) over the peak signal-to-noise ratio ($\sim$18) in \citet{Carral1999}. 
The emission peak of the knot would be located closer to the proposed star position than the Eastern masers (see Figure \ref{fig:G59_prop_map}), even considering its positional error. 
This fact may indicate that the H$_2$O masers trace a shock front preceding the radio jet. 
A similar spatial relation is observed in the radio jet and H$_2$O masers in IRAS 20126+4104 \citep{Moscadelli2005A&A, Hofner2007}. 
We note that the CH$_3$OH masers in IRAS 20126+4104 are excited in both the disc-like structure and low-velocity outflow, similar to the case of G59 \citep{Moscadelli2011}. 

\begin{figure}
	\includegraphics[width=\hsize]{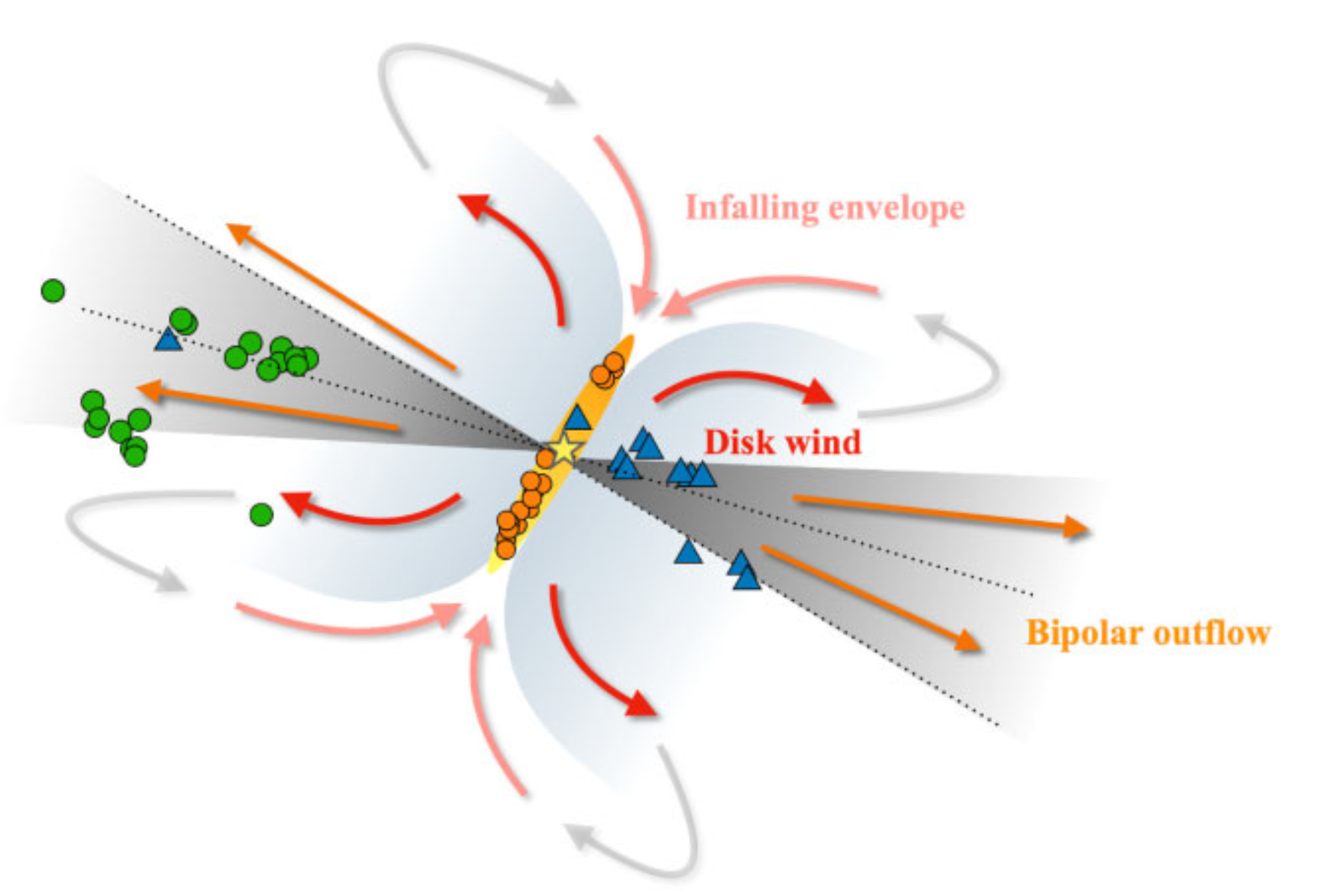}
     \caption{Schematic view of circumstellar structure inferred from our maser data. The markers are the same as Figure \ref{fig:G59_prop_map}.}
    \label{fig:G59_model}
\end{figure}

\section{Conclusions}
We have performed multi-epoch VLBI observation of 22 GHz H$_2$O and 6.7 GHz CH$_3$OH masers in the high-mass protostar G59.783+0.065, which shows a complex distribution of the class II CH$_3$OH maser. Figure \ref{fig:G59_model} summarises the schematic view of the circumstellar structure expected from our results. 
While H$_2$O masers trace a bipolar outflow along the NE-SW direction, the CH$_3$OH masers seem to have two different physical origins, as follows: 

\begin{enumerate}
\item The Eastern CH$_3$OH masers show a complex distribution and are closely located with the H$_2$O masers. Nevertheless, their proper motions are smaller than those of H$_2$O masers, and some CH$_3$OH masers have shown inward motions. This complex motion can be explained if these masers are excited at the boundary of the outflow and envelope. 

\item The Western CH$_3$OH masers show a linear distribution around the millimetre continuum peak. 
Their distribution is approximately perpendicular to the axis of the large scale outflow and possibly of a radio jet. The $PV$ diagram of the western CH$_3$OH masers suggests they might trace a rotating disc-like structure with possible evidence for expansion or infall.

\item However, the proper motions measured by JVN showed some discrepancy with respect to the proper motions expected for a disc-like structure which is either expanding or infalling.  
The Christmas-tree effect, changes of maser structures, and disc turbulence can be possible origins of observed randomness of the proper motion vectors. 

\item The complex distribution of the class II CH$_3$OH masers in G59 originates from simultaneous excitation by multiple physical structures, such as in the inner circumstellar envelope and outflow.  
More generally, we speculate that class II CH$_3$OH masers with complex distributions and different kinematics might trace gas belonging to both the inner circumstellar regions and the slow outflow regions at larger distances from the star. This hypothesis needs to be assessed by future statistical studies over a large sample of sources. 
\end{enumerate}

\section*{Acknowledgements}
The author thanks all the members in VERA and the Japanese VLBI network. 
This work was financially supported by the MEXT/JSPS KAKENHI Grant Numbers 15K17613, 19H05082, and 19H01937 (K.M.). 
This work is partially supported by the inter-university collaborative project `Japanese VLBI Network (JVN)' of NAOJ. 
This research has made use of data products from the Two Micron All Sky Survey, which is a joint project of the University of Massachusetts and the Infrared Processing and Analysis Center/California Institute of Technology, funded by the National Aeronautics and Space Administration and the National Science Foundation. This work is based, in part, on observations made with the Spitzer Space Telescope, which is operated by the Jet Propulsion Laboratory, California Institute of Technology under a contract with NASA. 

\section*{Data Availability}
The data underlying this article are archived by VERA and JVN. Please contact the corresponding author if one hope to access the data.  


\bibliographystyle{mnras}
\bibliography{Nakamura2023} 


\appendix
\section{Infrared view of G59} \label{IR_view}
\begin{landscape}
\begin{figure}
	\includegraphics*[width=0.95\hsize]{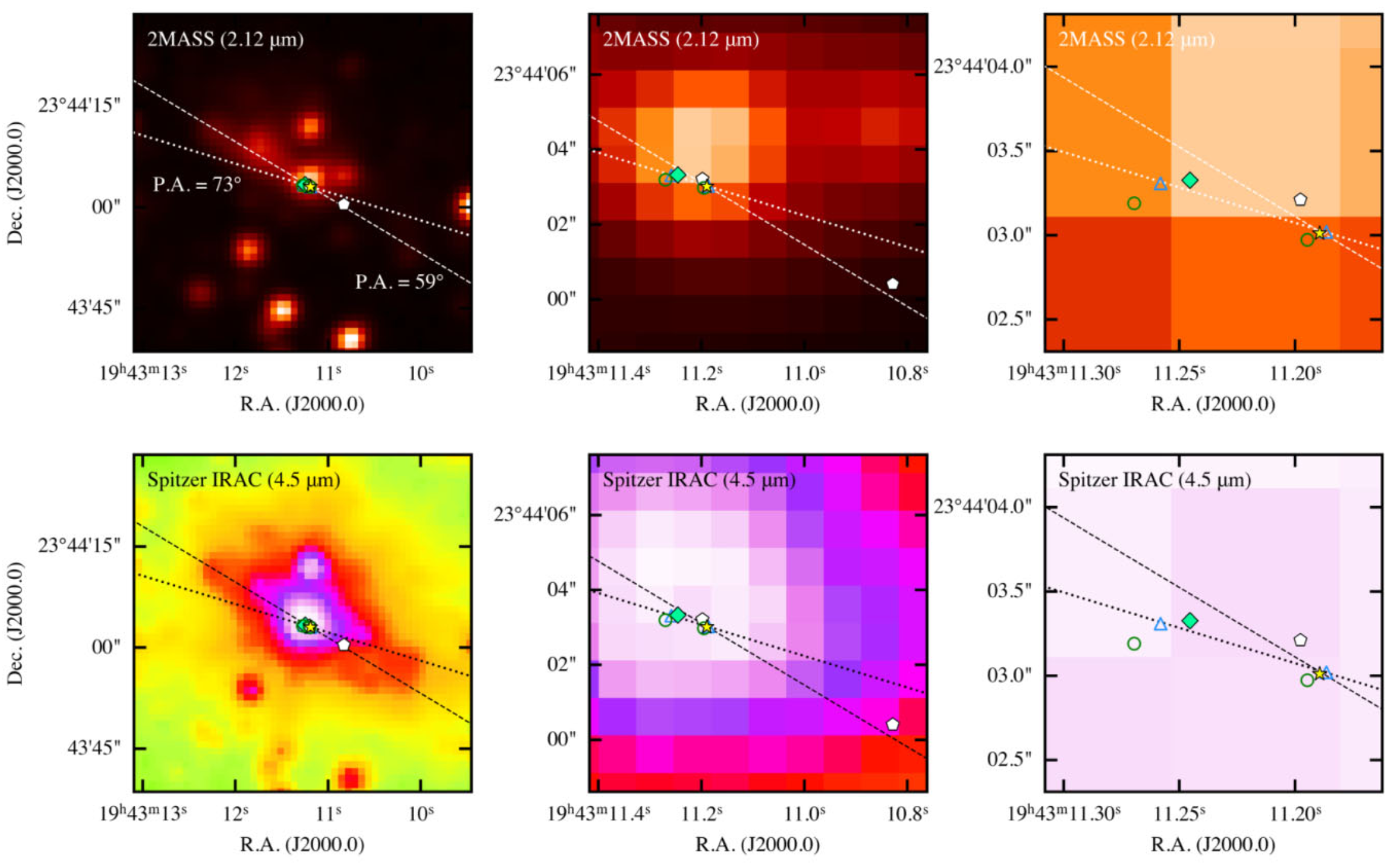}
    \caption{The upper panels present the near infrared image at 2.12 $\mu$m from 2MASS. The lower panels present the mid infrared image at 4.5 $\mu$m from Spitzer GLIMPSE survey. The open circles and squares indicates the CH$_3$OH and H$_2$O masers in this work, respectively. We only plotted two brightest maser features at the first epoch in the Western and Eastern maser clusters for both maser species. The white pentagons indicate the peak positions of the centimetre continuum source NE and SE \citep{Carral1999}. The green diamond shows the position of the another class II CH$_3$OH maser at 12 GHz measured by \citet{xu2009}. The yellow star indicates the position of the millimetre continuum source 13-s \citep{Rod2012}. The dotted and dashed lines show two possible position angles of the outflow, i.e., (1) along the bipolar distribution of H$_2$O masers and (2) perpendicular to the disc-like structure inferred from the linear alignment of CH$_3$OH masers, respectively.} 
    \label{fig:IR_view}
\end{figure}
\end{landscape}

Figure \ref{fig:IR_view} exhibits infrared view of G59 region from the Two Micron All Sky Survey \citep[2MASS:][]{Skrutskie2006AJ} and the Spitzer Galactic Legacy Infrared Mid-Plane Survey Extraordinaire \citep[GLIMPSE:][]{Benjamin2003PASP,Curchwell2009PASP}. 
We also showed several observational tracers in previous studies and this work. 

\section{elliptical fitting} \label{ellipse_fit}
\begin{figure}
	\includegraphics[width=\hsize]{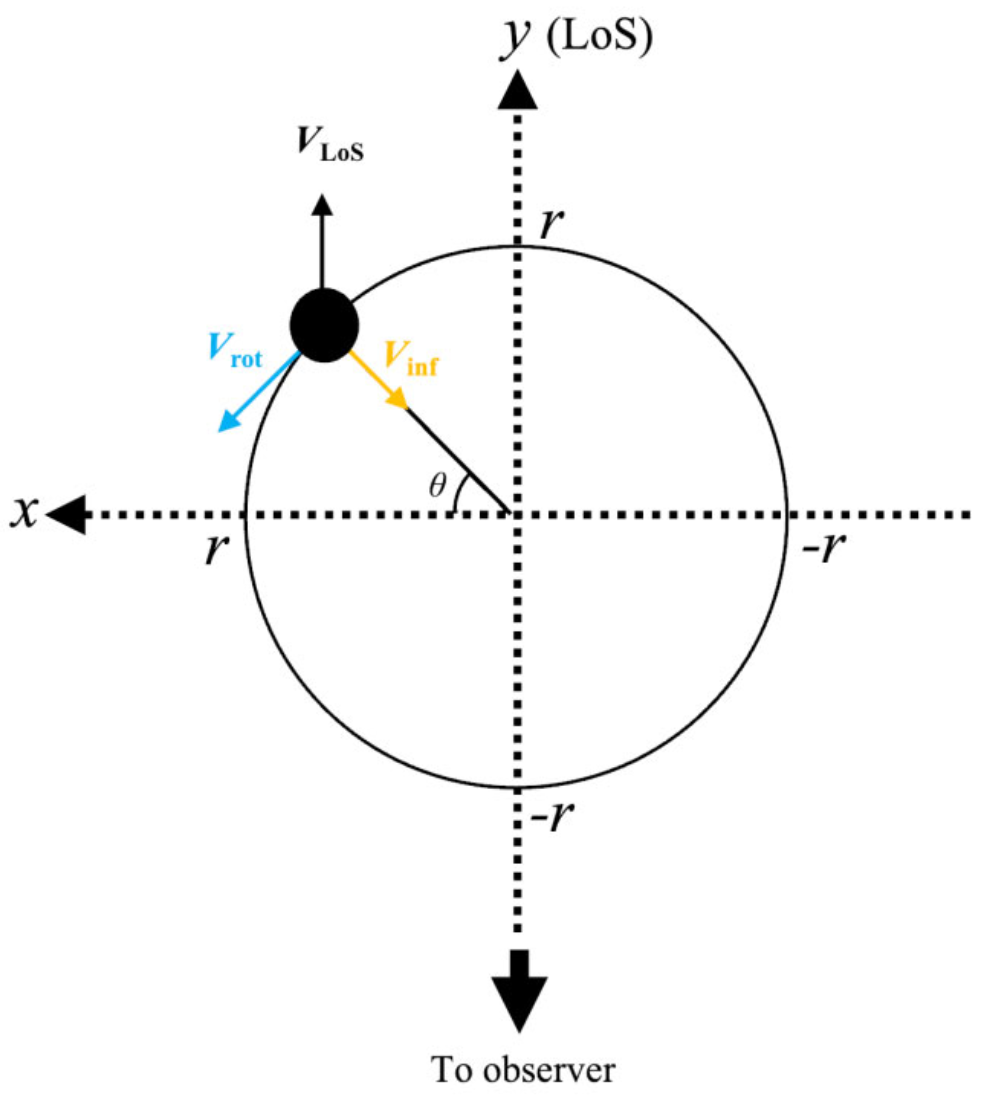}
    \caption{Parameter setup in the maser ring model}
    \label{fig:fit_set}
\end{figure}

Figure \ref{fig:fit_set} shows the setup of our maser ring model. 
An arbitrary point on the ring with a radius $r$ is expressed as $(x, y)$ = ($r\cos \theta, r\sin \theta$). Here, we set the $x$ axis perpendicular to the LoS as in the $PV$ diagram. The angle $\theta$ is measured clockwise from the $x$ axis to the LoS. 
We also define the rotating and infalling velocity of the ring as $V_{\rm rot}$ and $V_{\rm inf}$, respectively. 
The rotating velocity $V_{\rm rot}$ is measured counterclockwise. 
The expanding velocity ($V_{\rm exp}$) is given by $-V_{\rm inf}$. 
When we observe this disc edge-on, the LoS velocity $V_{\rm LoS}$ is as follows, 

\begin{eqnarray}
V_{\rm LoS} = V_\mathrm{exp} \sin \theta - V_\mathrm{rot} \cos \theta. 
\label{eq:vz}
\end{eqnarray}
The value of $V_{\rm LoS}$ is observationally determined by $V_\mathrm{LSR}-V_\mathrm{sys}$ at a given $x$ for each maser features. 
These $\cos\theta$ and $\sin\theta$ can be expressed by $r$, $V_{\rm LoS}$, $V_{\rm rot}$, and $V_{\rm exp}$ as follows, 
\begin{eqnarray}
\cos \theta = \frac{x}{r}, 
\label{eq:cos_sin1}
\end{eqnarray}

\begin{eqnarray}
\sin \theta = \frac{V_{\rm LoS}}{V_\mathrm{exp}} + \frac{x}{r} \cdot  \frac{V_\mathrm{rot}}{V_\mathrm{exp}}. 
\label{eq:cos_sin2}
\end{eqnarray}

We obtain an equation of the ellipse on the $PV$ diagram, substituting equation \ref{eq:cos_sin1} and \ref{eq:cos_sin2} into $\cos^2\theta+\sin^2\theta=1$. 
\begin{eqnarray}
\frac{1}{r^2} \left(1 + \frac{V_\mathrm{rot}^2}{V_\mathrm{exp}^2} \right) x^2 +  \frac{2V_\mathrm{rot}}{rV_\mathrm{exp}^2}xV_{\rm LoS} +\frac{1}{V_\mathrm{exp}^2}V_{\rm LoS}^2 =1
\label{eq:ellipse1}
\end{eqnarray}
The equation \ref{eq:ellipse1} can be simplified using parameters $A, B$, and $C$ as follows:  
\begin{eqnarray}
Ax^2 +  BxV_{\rm LoS} + CV_{\rm LoS}^2 = 1. 
\label{eq:ellipse2}
\end{eqnarray}
The parameters $A$, $B$, and $C$ were determined by the least square fitting on the observed $PV$ diagram. 
Table \ref{tab:pv_model} presents the best-fit parameters with 1-$\sigma$ errors. 
We finally obtain $r$, $V_{\rm rot}$, and $V_{\rm exp}$ (see Table \ref{tab:pv_parameter}), solving the following relations: 

\begin{eqnarray}
A &=& \frac{1}{r^2} \left(1+\frac{V_\mathrm{rot}^2}{V_\mathrm{exp}^2}\right) \\
B &=& \frac{2V_\mathrm{rot}}{rV_\mathrm{exp}^2}\\
C &=& \frac{1}{V_\mathrm{exp}^2}. 
\end{eqnarray}

\begin{table}
\centering
\caption{The best-fit parameters in the $PV$ model}
	\label{tab:pv_model}
    \begin{tabular}{lccc}  
        \hline\noalign{\vskip2pt} 
        \multirow{2}{*}{Parameters}  & \multirow{2}{*}{$A$} & \multirow{2}{*}{$B$}  & \multirow{2}{*}{$C$} \\  
              & & &  \\
        \multirow{2}{*}{}  & \multirow{2}{*}{[mas$^{-2}$]} & \multirow{2}{*}{[mas$^{-1}$ \ km$^{-1}$ \ s]}  & \multirow{2}{*}{[km$^{-2}$ \ s$^2$]} \\  
              & & &  \\          
        \hline 
         \multirow{2}{*}{Best-fit} &  \multirow{2}{*}{$5.918 \times 10^{-5}$} & \multirow{2}{*}{$7.711 \times 10^{-4}$}  & \multirow{2}{*}{$2.110\times 10^{-2}$} \\
         &                       &   &   \\
        Error ($1 \sigma$) &  $2.782\times10^{-6}$ & $1.292 \times 10^{-4}$ & $1.785 \times 10^{-3}$ \\
        \hline\noalign{\vskip2pt} 
    \end{tabular}
\end{table}

\section{All maser features detected}\label{all_feature_table}

\begin{table*}
    \centering
    \caption{All H$_2$O maser features detected}
    \label{tab:all_feature_table_w}
    \begin{tabular}{llrrrrrrrr}
    \hline
    Cluster$^{\rm a}$ & Name & $V_\mathrm{peak}$$^{\rm b}$ & $\Delta V$$^{\rm c}$ & $S_\mathrm{peak}$$^{\rm b}$ & $\Delta$RA$^{\rm d}$ & $\sigma$RA  & $\Delta$DEC$^{\rm d}$&  $\sigma$DEC & $N_{\rm spot}$$^{\rm e}$ \\ 
 &  & (km s$^{-1}$) &  (km s$^{-1}$) & (Jy) & (mas) & (mas) & (mas) & (mas) & \\ \hline
        \multicolumn{10}{c}{Epoch 1} \\ \hline
        E & w-1 & 25.15  & 0.84  & 2.05  & -154.090  & 0.017  & 116.476  & 0.017  & 5 \\ 
        E & w-2 & 27.47  & 2.11  & 7.46  & -156.833  & 0.302  & 118.451  & 0.487  & 11 \\ 
        E & w-3 & 28.73  & 1.90  & 4.64  & -157.942  & 0.197  & 119.277  & 0.146  & 10 \\ 
        W & w-4 & 21.99  & 1.69  & 11.01  & -1139.334  & 0.037  & -170.030  & 0.054  & 9 \\ 
        W & w-5 & 23.89  & 1.26  & 2.31  & -1139.992  & 0.071  & -170.255  & 0.019  & 7 \\ 
        W & w-6 & 17.36  & 4.21  & 27.03  & -1144.981  & 0.177  & -168.978  & 0.084  & 21 \\ 
        W & w-a & 4.09  & 2.53  & 3.26  & -1090.073  & 0.094  & -131.007  & 0.076  & 13 \\
        W & w-b & 20.73  & 0.63  & 1.07  & -1138.007  & 0.063  & -170.908  & 0.056  & 4 \\ 
        W & w-c & 21.57  & 4.42  & 9.51  & -1140.543  & 0.212  & -170.140  & 0.190  & 22 \\ \hline
        \multicolumn{10}{c}{Epoch 2} \\ \hline
        E & w-1 & 25.35  & 0.84  & 3.13  & -153.634  & 0.026  & 116.418  & 0.043  & 5 \\ 
        E & w-2 & 27.03  & 1.26  & 4.72  & -156.936  & 0.075  & 119.015  & 0.129  & 7 \\ 
        E & w-3 & 28.72  & 1.47  & 5.09  & -157.520  & 0.047  & 119.156  & 0.060  & 8 \\ 
        W & w-4 & 21.77  & 2.11  & 62.74  & -1139.329  & 0.351  & -170.088  & 0.045  & 11 \\ 
        W & w-5 & 24.08  & 1.05  & 1.83  & -1140.817  & 0.025  & -169.963  & 0.021  & 6 \\ 
        W & w-6 & 17.34  & 3.58  & 13.58  & -1144.983  & 0.178  & -168.945  & 0.101  & 18 \\ 
        W & w-b & 19.66  & 0.84  & 5.52  & -1137.141  & 0.029  & -170.779  & 0.007  & 5 \\ \hline
      \multicolumn{10}{c}{Epoch 3} \\ \hline
        E & w-1 & 25.39  & 0.84  & 3.01  & -153.076  & 0.012  & 116.317  & 0.037  & 5 \\ 
        E & w-2 & 27.28  & 0.63  & 0.65  & -156.506  & 0.039  & 118.903  & 0.044  & 4 \\ 
        E & w-3 & 28.55  & 1.69  & 5.24  & -156.979  & 0.061  & 119.082  & 0.032  & 9 \\ 
        W & w-4 & 21.59  & 1.69  & 11.38  & -1139.142  & 0.201  & -170.198  & 0.022  & 9 \\ 
        W & w-5 & 24.12  & 1.26  & 2.77  & -1140.647  & 0.025  & -170.042  & 0.024  & 7 \\
        W & w-6 & 17.38  & 1.47  & 3.85  & -1145.020  & 0.087  & -168.986  & 0.050  & 8 \\ 
        W & w-d & 11.90  & 0.63  & 1.27  & -1092.522  & 0.033  & -133.430  & 0.011  & 4 \\
        W & w-e & 19.28  & 1.47  & 1.65  & -1140.316  & 0.022  & -170.651  & 0.096  & 8 \\ 
        W & w-f & 19.28  & 1.69  & 1.17  & -1144.630  & 0.169  & -168.793  & 0.079  & 9 \\ \hline
      \multicolumn{10}{c}{Epoch 4$^{\rm f}$} \\ \hline
        W & w-4 & 21.55  & 0.63  & 4.62  & -1139.385  & 0.048  & -170.118  & 0.018  & 4 \\ \hline
\multicolumn{10}{l}{$^{\rm a}$ E: Eastern cluster, W: Western cluster. }\\
\multicolumn{10}{l}{$^{\rm b}$ $V_{LSR}$ and flux density of the peak maser spot in each maser feature. }\\
\multicolumn{10}{l}{$^{\rm c}$ Total velocity ranges of spectral channels where maser spots were detected. }\\
\multicolumn{10}{l}{$^{\rm d}$ Relative coordinate of maser features (see main text). }\\
\multicolumn{10}{l}{$^{\rm e}$ Number of maser spots consists of a maser feature.}\\
\multicolumn{10}{l}{$^{\rm f}$ Relative position was extrapolated by a measured proper motion (See main text). } \\
    \end{tabular}
\end{table*}

\begin{table*}
    \centering
    \caption{All CH$_3$OH maser features detected}
    \label{tab:all_feature_table_m}
    \begin{tabular}{llrrrrrrrr}
    \hline
    Cluster$^{\rm a}$ & Name & $V_\mathrm{peak}$$^{\rm b}$ & $\Delta V$$^{\rm c}$ & $S_\mathrm{peak}$$^{\rm b}$ & $\Delta$RA$^{\rm d}$ & $\sigma$RA  & $\Delta$DEC$^{\rm d}$&  $\sigma$DEC & $N_{\rm spot}$$^{\rm e}$ \\ 
 &  & (km s$^{-1}$) &  (km s$^{-1}$) & (Jy) & (mas) & (mas) & (mas) & (mas) & \\ \hline
\multicolumn{10}{c}{Epoch 1} \\ \hline
E & m-1  & 19.25  & 1.06  & 19.01  & -0.121  & 0.640  & -0.331  & 0.600  & 7 \\
E & m-2 & 17.32  & 0.36  & 1.55  & -20.445  & 0.140  & -6.197  & 0.189  & 3 \\
E & m-3 & 17.14  & 0.53  & 2.13  & -25.529  & 0.659  & -7.844  & 0.236  & 4 \\
E & m-4  & 15.56  & 0.35  & 1.79  & -165.977  & 0.168  & 135.722  & 0.157  & 3 \\
E & m-5 & 24.70  & 1.23  & 1.97  & -260.561  & 0.290  & 80.099  & 2.385  & 8 \\
E & m-6  & 27.15  & 1.05  & 17.41  & -313.786  & 0.594  & 83.363  & 1.220  & 7 \\
W & m-7 & 16.97  & 0.53  & 1.72  & -1003.316  & 0.513  & -268.261  & 0.368  & 4 \\
W & m-8  & 20.13  & 0.53  & 1.31  & -1009.536  & 0.162  & -273.207  & 0.053  & 4 \\
W & m-9  & 19.78  & 0.35  & 1.32  & -1018.565  & 0.023  & -263.700  & 0.471  & 3 \\
W & m-10 & 15.39  & 0.87  & 5.80  & -1026.592  & 0.161  & -217.186  & 0.288  & 6 \\
W & m-11 & 14.69  & 0.35  & 0.87  & -1029.893  & 0.553  & -213.909  & 0.542  & 3 \\
W & m-12 & 19.43  & 0.53  & 1.23  & -1114.213  & 0.214  & -83.831  & 0.184  & 4 \\
W & m-13 & 19.60  & 0.70  & 1.52  & -1125.450  & 1.538  & -71.219  & 1.108  & 5 \\
W & m-14 & 19.95  & 0.70  & 3.12  & -1136.994  & 0.595  & -64.560  & 0.754  & 5 \\
W & m-a & 15.39  & 0.53  & 0.69  & -1043.648  & 1.006  & -222.300  & 0.481  & 4 \\ \hline
\multicolumn{10}{c}{Epoch 2} \\ \hline
E & m-1 & 19.19  & 0.97  & 25.04  & -0.072  & 0.496  & -0.511  & 0.731  & 12 \\ 
E & m-2 & 17.35  & 0.17  & 0.79  & -21.124  & 0.310  & -7.721  & 0.111  & 3 \\
E & m-3 & 17.08  & 0.18  & 1.08  & -26.317  & 0.272  & -7.435  & 0.065  & 3 \\
E & m-4 & 15.50  & 0.18  & 0.57  & -166.481  & 0.274  & 135.527  & 0.173  & 3 \\
E & m-5 & 24.72  & 1.06  & 1.36  & -260.433  & 0.414  & 79.824  & 2.496  & 13 \\
E & m-6 & 27.09  & 0.79  & 8.02  & -314.063  & 0.722  & 82.968  & 1.253  & 10 \\
W & m-7 & 16.99  & 0.61  & 0.83  & -1004.203  & 1.353  & -268.477  & 0.472  & 8 \\
W & m-8 & 20.07  & 0.35  & 0.40  & -1009.762  & 0.256  & -272.921  & 0.377  & 5 \\
W & m-9 & 19.63  & 0.44  & 0.55  & -1018.672  & 0.455  & -263.771  & 0.762  & 6 \\
W & m-10 & 15.33  & 0.61  & 2.06  & -1026.708  & 0.730  & -216.963  & 0.474  & 12 \\
W & m-11 & 14.71  & 0.35  & 0.57  & -1030.730  & 0.350  & -213.594  & 0.292  & 5 \\
W & m-13 & 19.72  & 0.35  & 0.47  & -1126.333  & 0.584  & -70.865  & 0.392  & 5 \\
W & m-14 & 19.98  & 0.36  & 0.50  & -1136.803  & 0.410  & -64.817  & 0.095  & 5 \\ \hline
\multicolumn{10}{c}{Epoch 3} \\ \hline
E & m-1 & 19.20  & 0.70  & 22.89  & 0.006  & 0.667  & -0.275  & 0.479  & 9 \\
E & m-2 & 17.27  & 0.18  & 2.32  & -20.438  & 0.112  & -6.955  & 0.236  & 3 \\
E & m-3 & 17.09  & 0.17  & 3.91  & -25.968  & 0.254  & -7.897  & 0.104  & 3 \\
E & m-4 & 15.60  & 0.09  & 1.66  & -166.073  & 0.026  & 135.808  & 0.007  & 2 \\
E & m-5 & 24.73  & 0.88  & 3.78  & -260.645  & 0.354  & 79.804  & 2.332  & 11 \\
E & m-6 & 27.19  & 0.70  & 19.87  & -314.117  & 0.835  & 83.187  & 1.262  & 9 \\
W & m-7 & 16.92  & 0.09  & 2.19  & -1003.151  & 0.210  & -268.016  & 0.032  & 2 \\
W & m-8 & 20.17  & 0.17  & 2.70  & -1010.044  & 0.034  & -272.743  & 0.286  & 3 \\
W & m-10 & 15.34  & 0.62  & 7.05  & -1026.692  & 0.319  & -217.202  & 0.322  & 8 \\
W & m-12 & 19.55  & 0.17  & 3.33  & -1115.891  & 1.054  & -83.102  & 0.910  & 3 \\
W & m-13 & 19.64  & - & 2.24  & -1124.373  & 0.349  & -72.566  & 0.398  & (1)$^{\rm f}$ \\
W & m-14 & 19.99  & 0.62  & 19.99  & -1136.953  & 0.739  & -64.997  & 0.743  & 8 \\ \hline
\multicolumn{10}{l}{$^{\rm a}$ E: Eastern cluster, W: Western cluster. }\\
\multicolumn{10}{l}{$^{\rm b}$ $V_{LSR}$ and flux density of the peak maser spot in each maser feature. }\\
\multicolumn{10}{l}{$^{\rm c}$ Total velocity ranges of spectral channels where maser spots were detected. }\\
\multicolumn{10}{l}{$^{\rm d}$ Relative coordinate of maser features (see main text). }\\
\multicolumn{10}{l}{$^{\rm e}$ Number of maser spots consists of a maser feature.}\\
\multicolumn{10}{l}{$^{\rm f}$ Since only a single spot was detected, the m-13 feature at third epoch was not used for any analysis.} \\
\end{tabular}
\end{table*}

\begin{table*}
    \centering
    \contcaption{All CH$_3$OH maser features detected}
    \begin{tabular}{llrrrrrrrr}
    \hline
    Cluster$^{\rm a}$ & Name & $V_\mathrm{peak}$$^{\rm b}$ & $\Delta V$$^{\rm c}$ & $S_\mathrm{peak}$$^{\rm b}$ & $\Delta$RA$^{\rm d}$ & $\sigma$RA  & $\Delta$DEC$^{\rm d}$&  $\sigma$DEC & $N_{\rm spot}$$^{\rm e}$ \\ 
 &  & (km s$^{-1}$) &  (km s$^{-1}$) & (Jy) & (mas) & (mas) & (mas) & (mas) & \\ \hline
\multicolumn{10}{c}{Epoch 4} \\ \hline
E & m-1 & 19.18  & 0.74  & 42.43  & -0.176  & 0.718  & -0.394  & 0.556  & 19 \\
E & m-2 & 17.25  & 0.31  & 4.10  & -21.730  & 0.989  & -6.778  & 0.468  & 8 \\
E & m-3 & 17.12  & 0.18  & 10.14  & -26.957  & 0.111  & -7.429  & 0.158  & 5 \\
E & m-4 & 15.57  & 0.08  & 2.98  & -167.336  & 0.249  & 136.348  & 0.344  & 3 \\
E & m-5 & 24.58  & 1.10  & 11.14  & -261.710  & 0.562  & 80.809  & 2.152  & 26 \\
E & m-6 & 27.17  & 0.74  & 40.29  & -315.750  & 0.845  & 83.525  & 1.575  & 18 \\
W & m-7 & 17.11  & 0.13  & 2.15  & -1004.387  & 0.183  & -267.125  & 0.298  & 2 \\
W & m-8 & 20.18  & 0.13  & 3.31  & -1010.827  & 0.207  & -272.848  & 0.295  & 4 \\
W & m-10 & 15.32  & 0.53  & 8.86  & -1028.277  & 0.552  & -216.423  & 0.430  & 13 \\
W & m-11 & 14.92  & 0.53  & 7.15  & -1032.010  & 0.654  & -213.249  & 0.655  & 13 \\
W & m-12 & 19.49  & 0.18  & 5.66  & -1116.469  & 1.321  & -83.073  & 1.210  & 5 \\
W & m-13 & 19.71  & 0.49  & 12.23  & -1125.886  & 1.013  & -72.143  & 0.790  & 12 \\
W & m-14 & 19.97  & 0.22  & 3.98  & -1138.504  & 0.465  & -64.601  & 0.318  & 6 \\ \hline
\multicolumn{10}{c}{Epoch 5} \\ \hline
E & m-1 & 19.19  & 0.75  & 88.39  & -0.144  & 0.619  & -0.399  & 0.636  & 19 \\
E & m-2 & 17.31  & 0.22  & 9.87  & -21.531  & 0.301  & -6.565  & 0.351  & 6 \\
E & m-3 & 17.09  & 0.31  & 22.27  & -26.522  & 0.823  & -7.455  & 0.295  & 8 \\
E & m-4 & 15.55  & 0.13  & 6.00  & -167.614  & 0.360  & 136.710  & 0.160  & 4 \\
E & m-5 & 24.16  & 1.05  & 21.19  & -261.895  & 0.578  & 80.702  & 2.174  & 27 \\
E & m-6 & 27.18  & 0.75  & 81.20  & -315.782  & 0.920  & 83.695  & 1.444  & 18 \\
W & m-9 & 19.81  & 0.53  & 28.95  & -1020.642  & 0.503  & -263.638  & 0.449  & 13 \\
W & m-10 & 15.38  & 0.57  & 18.44  & -1028.179  & 0.660  & -216.453  & 0.359  & 14 \\
W & m-11 & 14.89  & 0.48  & 14.11  & -1031.831  & 0.735  & -213.335  & 0.718  & 12 \\
W & m-12 & 19.50  & 0.22  & 14.83  & -1116.180  & 1.321  & -83.266  & 1.204  & 6 \\
W & m-13 & 19.63  & 0.26  & 27.58  & -1125.871  & 1.089  & -72.131  & 0.785  & 12 \\
W & m-14 & 19.94  & 0.26  & 14.19  & -1138.406  & 0.269  & -64.681  & 0.192  & 7 \\
W & m-b  & 15.55  & 0.09  & 10.15  & -1013.129  & 0.168  & -192.151  & 0.058  & 3 \\ \hline
\multicolumn{10}{c}{Epoch 6} \\ \hline
E & m-1 & 19.22  & 0.83  & 45.71  & -0.319  & 0.704  & -0.499  & 0.620  & 21  \\
E & m-2 & 17.29  & 0.26  & 5.42  & -21.115  & 0.421  & -6.058  & 0.190  & 7  \\
E & m-3 & 17.11  & 0.35  & 12.11  & -26.864  & 0.888  & -7.486  & 0.264  & 9  \\
E & m-4 & 15.66  & -0.18  & 15.66  & -167.779  & 0.251  & 137.019  & 0.189  & 5  \\
E & m-5 & 24.44  & 1.14  & 17.51  & -262.356  & 2.204  & 80.705  & 2.204  & 32  \\
E & m-6 & 27.21  & 0.75  & 41.78  & -316.367  & 0.825  & 83.880  & 1.463  & 18  \\
W & m-9 & 19.83  & 0.48  & 10.54  & -1021.039  & 0.437  & -263.347  & 0.578  & 13  \\
W & m-10 & 15.31  & 0.57  & 15.31  & -1028.506  & 0.483  & -216.324  & 0.326  & 14  \\
W & m-11 & 14.87  & 0.66  & 9.03  & -1032.338  & 0.596  & -213.122  & 0.572  & 16  \\
W & m-13 & 19.70  & 0.57  & 13.24  & -1126.276  & 1.092  & -72.138  & 0.780  & 14  \\
W & m-14 & 20.01  & 0.04  & 2.61  & -1139.023  & 0.311  & -64.488  & 0.404  & 2  \\
E & m-c & 20.18  & 0.13  & 2.89  & 78.740  & 0.312  & 185.986  & 0.186  & 4  \\ \hline
\multicolumn{10}{l}{$^{\rm a}$ E: Eastern cluster, W: Western cluster. }\\
\multicolumn{10}{l}{$^{\rm b}$ $V_{LSR}$ and flux density of the peak maser spot in each maser feature. }\\
\multicolumn{10}{l}{$^{\rm c}$ Total velocity ranges of spectral channels where maser spots were detected. }\\
\multicolumn{10}{l}{$^{\rm d}$ Relative coordinate of maser features (see main text). }\\
\multicolumn{10}{l}{$^{\rm e}$ Number of maser spots consists of a maser feature.}\\
\end{tabular}
\end{table*}

\bsp	
\label{lastpage}
\end{document}